\documentclass{aa}  

\usepackage{graphicx}
\usepackage{tabularx}
\usepackage{txfonts}
\usepackage{xcolor}
\usepackage{booktabs} 
\usepackage{float} 

\begin{document} 

   \title{AGN with massive black holes have closer galactic neighbors: }
   \subtitle{$k-$Nearest-Neighbor Statistics of an unbiased sample of AGN at $z\sim 0.03$}

   \author{A. Mhatre
          \inst{1},
          M. C. Powell \inst{2,1}\thanks{\email{mpowell@aip.de}},
          S. Yuan \inst{1},
          S. W. Allen \inst{1},  T. Caglar \inst{3,4}
          M. Koss \inst{5},  I. del Moral-Castro \inst{6}, K. Oh \inst{7}, A. Peca \inst{5,8}, C. Ricci \inst{9,10},
          F. Ricci \inst{11,12},  A. Rojas \inst{13}, M. Signorini \inst{14,15}
          }

   \institute{Kavli Institute for Particle Astrophysics and Cosmology, Stanford University, 452 Lomita Mall, Stanford, CA 94305, USA
         \and
         Leibniz-Institut fur Astrophysik Potsdam (AIP), An der Sternwarte 16, D-14482 Potsdam, Germany
         \and
        George P. and Cynthia Woods Mitchell Institute for Fundamental Physics and Astronomy, Texas A\&M University, College Station, TX, 77845, USA
         \and
         Leiden Observatory, PO Box 9513, 2300 RA Leiden, The Netherlands
         \and
         Eureka Scientific, 2452 Delmer Street Suite 100, Oakland, CA 94602-3017, USA
        \and
         Instituto de Astrofísica, Facultad de Física, Pontificia Universidad Católica de Chile, Campus San Joaquín, Av. Vicuña Mackenna 4860, Macul, Santiago, Chile, 7820436
         \and
        Korea Astronomy and Space Science Institute, 776, Daedeokdae-ro, Yuseong-gu, Daejeon 34055, Republic of Korea
        \and
         Yale Center for Astronomy \& Astrophysics and Department of Physics, Yale University, P.O. Box 208120, New Haven, CT 06520-8120, USA
         \and
         Nucleo de Astronomıa de la Facultad de Ingenieria, Universidad Diego Portales, Av. Ejercito Libertador 441, Santiago, Chile
         \and
         Kavli Institute for Astronomy and Astrophysics, Peking University, Beijing 100871, People’s Republic of China
         \and
         Dipartimento di Matematica e Fisica, Universit`a degli Studi Roma Tre, Via della Vasca Navale 84, I-00146, Roma, Italy
         \and
         INAF - Osservatorio Astronomico di Roma, via Frascati 33, 00040 Monteporzio Catone, Italy
         \and
        Departamento de F\'isica, Universidad T\'ecnica Federico Santa Mar\'ia, Vicu\~{n}a Mackenna 3939, San Joaqu\'in, Santiago de Chile, Chile
         \and
         INAF– Osservatorio Astrofisico di Arcetri, Largo Enrico Fermi 5, I-50125 Firenze, Italy
         \and
         Dipartimento di Matematica e Fisica, Univeristà di Roma 3, Via della Vasca Navale, 84, 00146 Roma RM
}

   \date{}

  \abstract
{The large-scale environments of active galactic nuclei (AGN) reveal important information on the growth and evolution of supermassive black holes (SMBHs). Previous AGN clustering measurements using 2-point correlation functions have hinted that AGN with massive black holes preferentially reside in denser cosmic regions than AGN with less-massive SMBHs. At the same time, little to no dependence on the accretion rate is found. 
However, the significance of such trends have been limited. }
{Here we present $k^{th}$-nearest-neighbor ($k$NN) statistics of 2MASS galaxies around AGN from the Swift/BAT AGN Spectroscopic survey. These statistics have been shown to contribute additional higher-order clustering information on the cosmic density field. }
{By calculating the distances to the nearest 7 galaxy neighbors in angular separation to each AGN within two redshift ranges ($0.01<z<0.03$ and $0.03<z<0.06$), we compare their cumulative distribution functions to that of a randomly distributed sample to show the sensitivity of this method to the clustering of AGN. We also split the AGN into bins of bolometric luminosity, black hole mass, and Eddington ratio (while controlling for redshift) to search for trends between $k$NN statistics and fundamental AGN properties.}
{We find that AGN with massive SMBHs have significantly closer neighbors than AGN with less-massive SMBHs (at the 99.98\% confidence level), especially in our lower redshift range. We find less significant trends with luminosity or Eddington ratio. 
By comparing our results to empirical SMBH-galaxy-halo models implemented in N-body simulations, we show that small-scale $k$NN trends with black hole mass may go beyond stellar mass dependencies.  }
{This suggests that massive SMBHs in the local universe reside in more massive dark matter halos and denser regions of the cosmic web, which may indicate that environment is important for the growth of SMBHs, bolstering previous conclusions using correlation functions. }


\authorrunning{A. Mhatre, M. C. Powell, et al.}
\titlerunning{AGN with massive supermassive black holes have closer galactic neighbors}

\maketitle
%

\section{Introduction}

Supermassive black holes (SMBHs) are ubiquitously found at the centers of massive galaxies. Phases of their mass growth occur when they actively accrete matter, which we observe as active galactic nuclei (AGN). Studies of nearby galaxies and AGN over the last few decades have shown that there are several correlations between SMBHs and properties of their host galaxies, which suggest that their growth and evolution are linked in some way \citep[e.g.,][]{Kormendy:2013}. At the same time, galaxy properties--like mass, morphology, star formation rate, and size--have been found to correlate with their larger scale dark matter (DM) halo environments, indicating that environment and the dark matter gravitational potential wells impact how galaxies evolve over cosmic time \cite[e.g.,][]{coil:2017,Behroozi:2019,ghosh:2024}. 
Still, it remains unclear which physical processes drive the growth and coevolution between SMBHs, galaxies, and their DM halos, as well as the role of the larger-scale environment. 
Characterizing the trends between SMBHs and their host dark matter halos can provide clues to these processes.

The spatial clustering of AGN probes the large-scale environments of growing SMBHs. 
Two point statistics are a well-established method of measuring clustering, and have provided clues for the link between AGN and large-scale structure across a range of spatial scales. For example, detected AGN seem to reside in halos that have masses characteristic of galaxy group environments \citep[e.g.,][]{Cappelluti:2012}. Additionally, there seems to be little or no dependence between the AGN clustering amplitude and luminosity \citep{Krumpe:2012,Allevato:2011,Powell:2020}, which would imply that the accretion rate does not depend strongly on the extragalactic environment. However, previous studies from X-ray surveys have shown that there may be trends with black hole mass ($M_{\rm BH}$); AGN with more massive SMBHs are typically found to be more clustered than less-massive SMBHs \citep{Shirasaki:2016,Powell:2018,Powell:2022,Krumpe:2015,Krumpe:2023}. However, the relatively low number densities of AGN samples have so far limited the significance of such trends. An additional complication is the many AGN selection methods that each have their own biases against the underlying population of accreting supermassive black holes; AGN clustering has been found to strongly depend on the selection method \citep[e.g.,][]{Mendez:2016,Powell:2024}. 
Previous studies have also explored potential clustering differences between obscured and unobscured AGN, often finding conflicting results \citep[e.g.,][]{Powell:2018,DiPompeo:2017,Viitanen:2021}; selection biases are likely causing some of the inconsistencies.
Additional sources of uncertainties and degeneracies are the conversion between the large-scale clustering amplitude and halo mass \citep{DeGraf:2017,Oogi:2020,Aird:2021}, and whether or not underlying host galaxy population is the main driver of AGN clustering \citep{Mendez:2016,Allevato:2019,Krishnan:2020,Powell:2022}.  

The Swift/BAT Spectroscopic Survey\footnote{\url{https://bass-survey.com}} is one of the largest and most unbiased samples of local AGN ($z<0.1$; BASS DR2; \citealt{Koss:2017,Koss:2022b}). Its hard X-ray selection (14-195 keV) is sensitive to unobscured and obscured sources (nearly unbiased to obscuring column densities up to $N_{H}=10^{24}$ cm$^{-2}$; e.g., \citealt{Ricci:2015,Ananna:2022}), and the abundance of multiwavelength follow-up, including optical spectroscopy and soft X-ray observations \citep{Ricci:2017B}, has allowed for the characterization of the vast majority of BASS AGN. The survey also overlaps with the Two Micron All-
Sky Survey (2MASS) galaxy survey \citep{Huchra:2012}, enabling powerful AGN-galaxy cross-correlation measurements that boost clustering statistics \citep{Koss:2010,Cappelluti:2010,Powell:2018}. Previous AGN clustering measurements with this sample via the 2-point correlation function have found hints that AGN with more massive SMBHs cluster stronger than AGN with less-massive SMBHs on scales of the 1-halo term (e.g., $<1~h^{-1}$Mpc; \citealt{Powell:2022}). While the differences were only at a level of $2.3\sigma$, it was shown in \citealt{Powell:2022} that such trends can constrain the underlying relationship between SMBH mass and the host dark matter halo mass, which would provide a powerful benchmark for SMBH growth implemented in hydrodynamic simulations. Therefore, characterizing AGN clustering trends at a higher significance would be an invaluable constraint for AGN fueling and feedback processes.
While AGN 2-point correlation function measurements will improve over time with future larger surveys like \textit{Euclid} and eROSITA/SDSS-V/4MOST, we can alternatively use additional probes of the cosmic density field of the current samples to better understand the AGN-halo connection.

$k$-Nearest Neighbor ($k$NN) Cumulative Distribution Functions have proven to be a powerful complementary probe of galaxy clustering \citep[e.g.,][]{Banerjee_2020,Wang_2022}. $k$NN statistics comprise the distributions of distances between a set of points and their $k^{th}$ nearest-neighbors, where many different values of $k$ are explored. With sufficient values of $k$, these statistics are sensitive to higher-order correlation functions on nonlinear physical scales, which 2-point functions are not \citep{Yuan_2023}. This provides additional information on the clustering statistics of galaxies. While $k$NN statistics have been primarily applied to galaxy and galaxy clusters \citep[e.g.,][]{Banerjee_2020,Wang_2022, Yuan_2023}, this method has not yet been fully explored with AGN samples for studying the SMBH-halo connection.

Previous studies of galactic neighbors around AGN have typically used only 1-3 values for $k$, e.g., calculating only the 1st or 5th nearest neighbor distances, which have limited the full potential of such statistics \citep[e.g.,][]{Silverman:2009,Perez:2022}. 
Additionally, trends with the fundamental AGN parameters (e.g., mass and accretion rate) have been unexplored with $k$NN statistics, especially using a large and unbiased sample.

Here we measure the $k$NN statistics for the BASS DR2 hard X-ray-selected AGN sample to investigate trends between the fundamental AGN properties and large-scale structure. We measure the angular distances to the first seven neighbors to characterize their environments on small ($<1 ~h^{-1}$Mpc) and larger ($1-4h^{-1}$Mpc) scales in two redshift bins, and we test whether or not there are significant trends with luminosity, black hole mass, and Eddington ratio. We also briefly examine trends with obscuring column density to eliminate possible degeneracies in black hole mass differences. The paper is organized as follows: we first describe the AGN and galaxy samples used in our analyses (Section \ref{sec:data}), including selection criteria and estimations of black hole masses and luminosities. Next, we explain the methodology of $k$NN statistics, how we apply it to our data, and our metrics to verify our results (Section \ref{sec:methods}). In Section \ref{sec:results},  we present the findings of our $k$NN analysis for black hole mass, luminosity, and accretion rate, and in Section \ref{sec:mods} we compare our results to simple empirical models implemented in an N-body simulation. In Sections \ref{sec:discussion} and \ref{sec:summary} we discuss our results and summarize our findings.

\begin{figure}
  \centering
\includegraphics[width=0.49\textwidth]{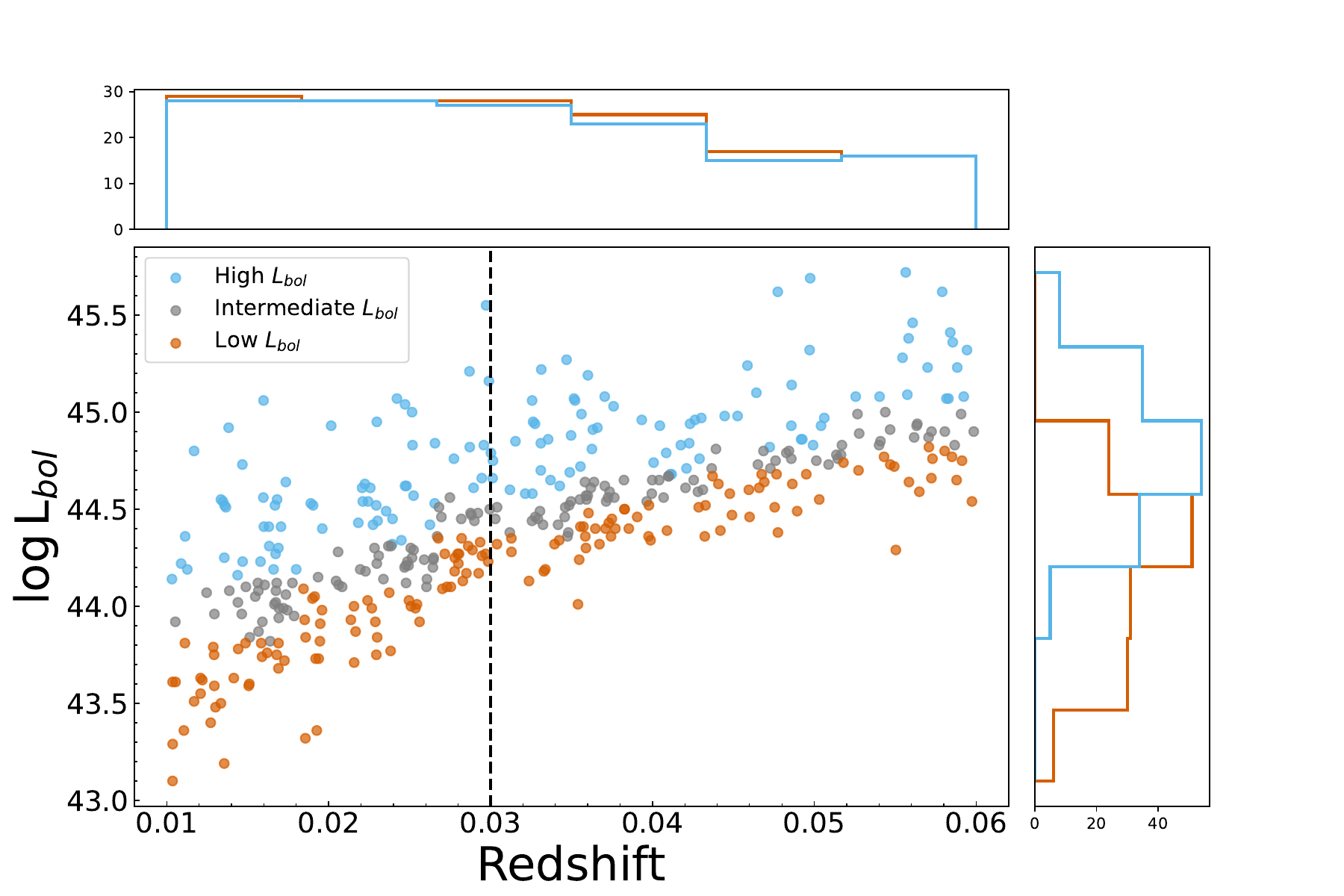}
\includegraphics[width=0.49\textwidth]{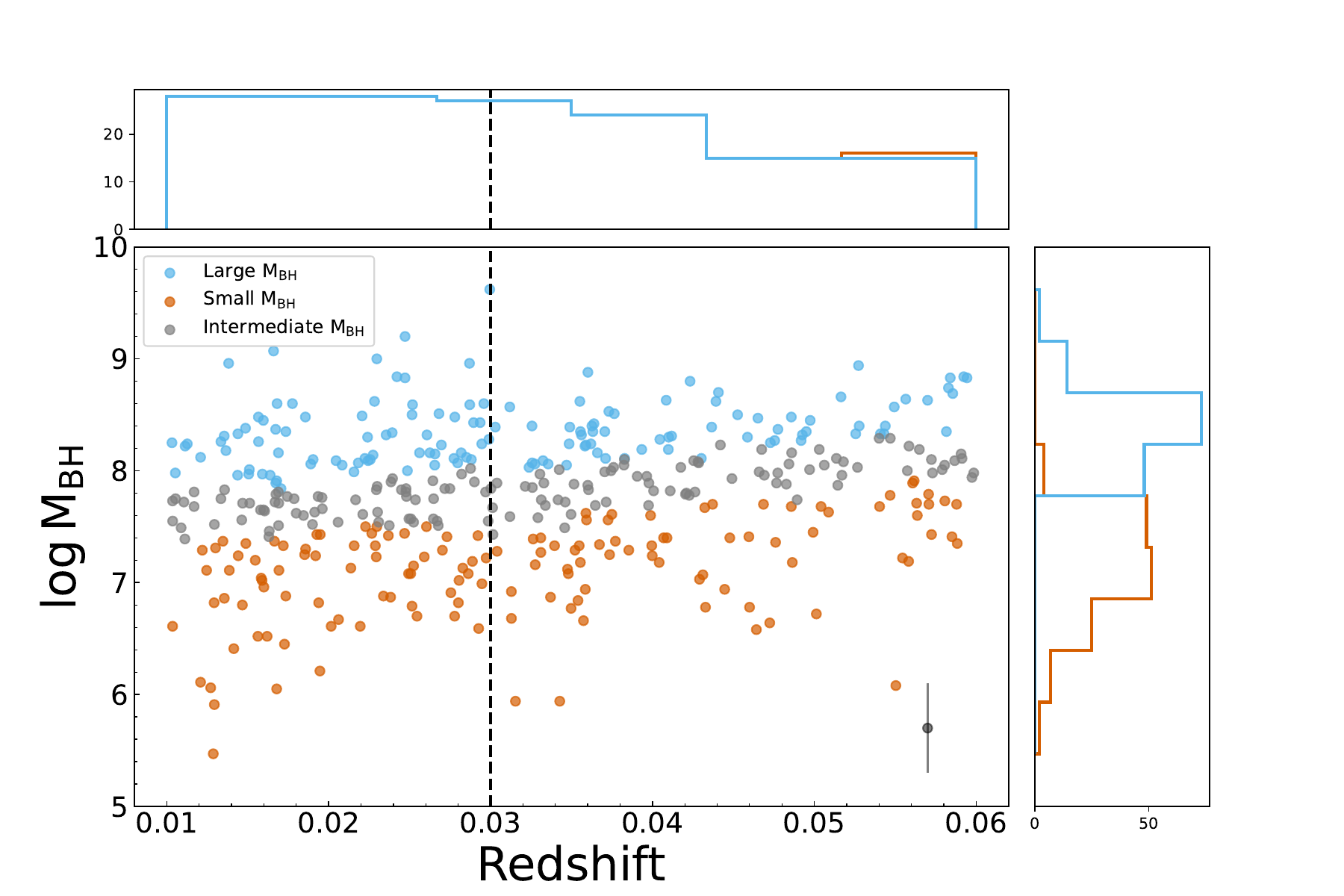}
\includegraphics[width=0.49\textwidth]{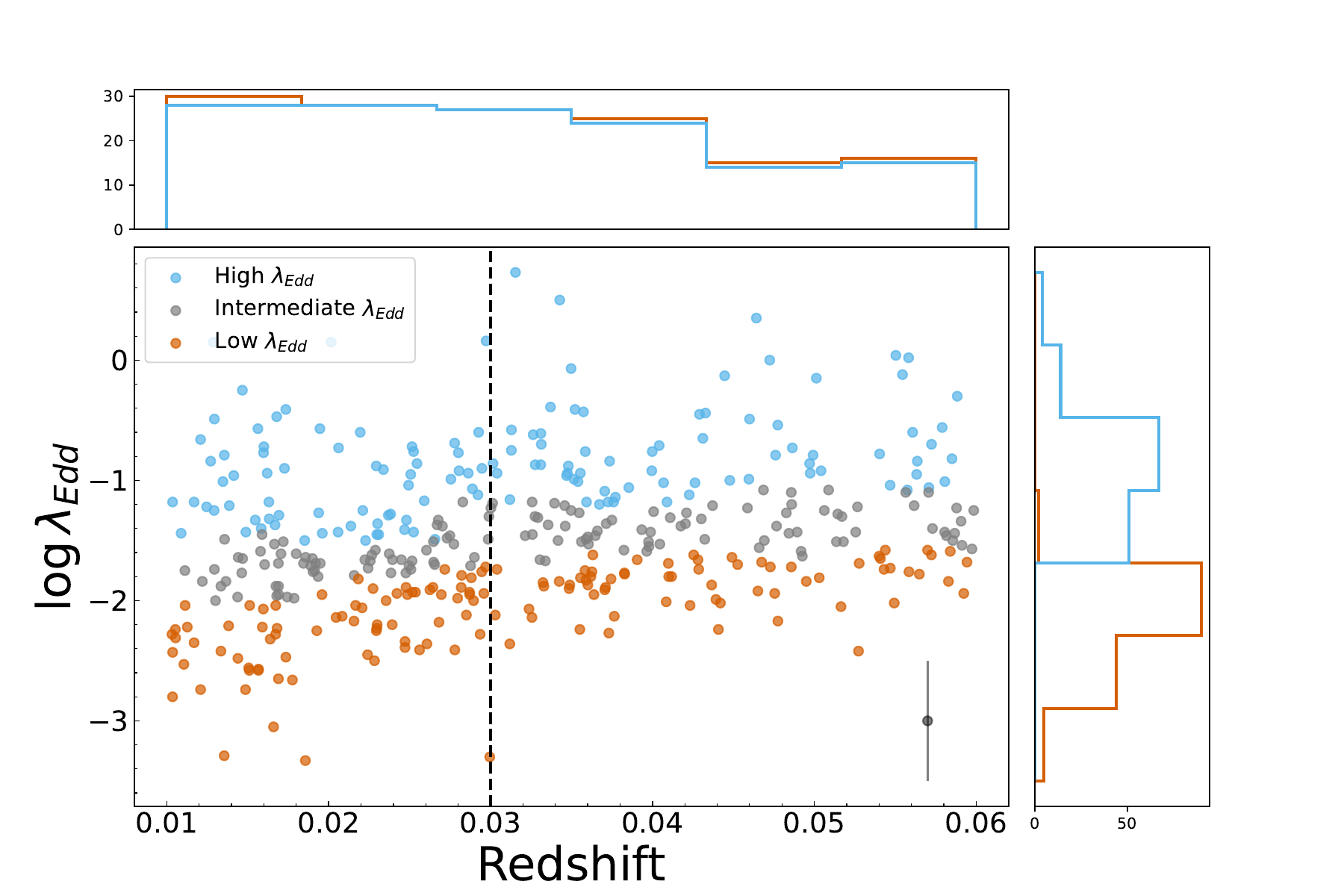}

  \caption{AGN parameters (luminosity, black hole mass, and Eddington ratio) vs. redshift for the BASS AGN sample used in our $k$NN analysis. We categorized the parameter bins into large (sky blue) and small (orange) subsamples to have similar redshift distributions. For each plot, the top histogram shows the redshift distribution of the two bins, and the rightward histogram shows the parameter distribution. Luminosity bins are shown at the top, black hole mass bins are in the middle, and Eddington ratio bins are at the bottom. We show the typical parameter uncertainties in the bottom right.}
  \label{fig:mbh-luminosity-distribution}
\end{figure}


\section{Data}
\label{sec:data}
\subsection{X-ray-selected AGN Sample}
The AGN sample is taken from the second data release of the Swift/BAT AGN Spectroscopic Survey (BASS DR2; \citealt{Koss:2022a,Koss:2022b}), which is drawn from the 70-month catalog of detections from the hard X-ray BAT detector aboard the Neil Gehrels Swift observatory (\citealt{Baumgartner:2013}; sensitive from $14-195$ keV). Excluding beamed and lensed AGN \citep{Marcotulli:2022}, the full sample comprises 858 AGN. We selected objects within the redshift range $0.01<z<0.06$ outside of the galactic plane ($|b|>8^{\circ}$), totaling 419 AGN.

Optical spectroscopy was obtained for each AGN, enabling secure spectroscopic redshift measurements and black hole mass estimates for 98\% of the sample. For the Type 1 AGN with broad ($>1000$ km s$^{-1}$) emission lines, black hole masses were calculated via the full width at half maximum (FWHM) of H$\alpha$, H$\beta$, Mg II, and/or C IV \citep{Mejia-Restrepo:2022}. For Type 2 AGN without broad Balmer lines, bulge velocity dispersions ($\sigma_{*}$) were measured by the absorption features in the optical spectra (the Ca H+K, Mg I, and/or the Ca II triplet); the $M_{\rm BH}-\sigma_{*}$ from \cite{Kormendy:2013} was assumed for $M_{\rm BH}$ \citep{Koss:2022b}. Of the 419 AGN, 412 have black hole mass measurements; 225 from stellar velocity dispersion measurements, 157 from broad line measurements, and 30 from the literature via reverberation mapping or dynamical measurements.
For this analysis, we use the best $M_{\rm BH}$ measurement available for each AGN. However, a bias has been found between the two main $M_{\rm BH}$ measurement techniques, based on comparing Type 1 $M_{\rm BH}$ estimates obtained with $\sigma_{*}$ to their broad line measurements \citep{Caglar:2020,caglar:2023}. We therefore repeated our analysis using consistently-measured $M_{BH}$ values and ensured that this bias is not significantly impacting our main results, which we present in Appendix \ref{sec:appendixC}.

In addition to optical spectroscopy, each AGN was followed up with soft X-ray observations by Swift, Chandra, and/or XMM-Newton. This has enabled robust measurements of the absorbed column densities ($N_{\rm H}$) and the intrinsic X-ray luminosities \citep{Ricci:2017B}. Column densities range from the unabsorbed to Compton-thick level ($\log~N_{\rm H}=25$ [$cm^{-2}$]). The bolometric luminosities were estimated based on the intrinsic ultra-hard luminosities ($L_{14-195}$) via $L_{bol}=8\times L_{14-195}$. 

In this work, we calculate the nearest-neighbor statistics in bins of luminosity ($L_{\rm bol}$), black hole mass ($M_{\rm BH}$), and Eddington ratio ($\lambda_{\rm Edd}\equiv L_{\rm bol}/L_{\rm Edd}$, where $L_{\rm Edd}$ is the Eddington luminosity). We controlled for redshift for each binning scheme so that the same cosmic volumes were probed and the projection effects were the same for each bin. This was done via the following methodology: we first divided the sample into six equally spaced redshift intervals from $z=0.01$ to $z=0.06$; then, within each redshift interval, AGNs were further subdivided based on the AGN parameter ($L_{\rm bol}$, $M_{\rm BH}$, or $\lambda_{Edd}$) into three tertiles. The upper bin was defined as the upper 33\% tertile; the lower bin was defined as the lower 33\% tertile. We deliberately chose to exclude the middle tertile to minimize overlap and potential contamination in each bin due to measurement uncertainties. For black hole mass estimates, for example, uncertainties range from 0.3 to 0.5 dex \cite[e.g.,][]{Koss:2022a,Mejia-Restrepo:2022}. 
The number of AGN in each subsample is given in Table \ref{tab:tab1}, along with the median redshifts, luminosities, black hole masses, and Eddington ratios. We additionally defined two redshift ranges to perform a tomographic analysis and probe projected spatial scales from $0.1$ to $4~h^{-1}$Mpc: $0.01<z<0.03$ and $0.03<z<0.06$. We show the distributions of each parameter vs. redshift for each parameter bin in Fig. \ref{fig:mbh-luminosity-distribution}.

\begin{table*}[]
    \centering
\begin{tabular}{cc|cccccc}
\hline
$z-$range & Subsample                     &   \# of Objects &   Median BH Mass &   Median $L_{\rm{bol}}$ &   Median $z$ & Median $\lambda_{\rm Edd}$ \\
 &                    &   &    [dex $M_{\odot}$]&  [dex erg/s] &   & [dex]\\
\hline
\hline

& High-$L$ &             63 &            8.09  &         44.54 &             0.022 & -1.67\\
& Low-$L$   &             67 &            7.43  &         43.84   &             0.020 & -1.69\\
 $0.01<z<0.03$ & Large M$_{\rm BH}$       &             70 &            8.24  &         44.39 &             0.022 & -2.10\\
& Small M$_{\rm BH}$      &             69 &           7.08 &         44.09 &             0.020 & -1.12\\
& High-$\lambda_{Edd}$ &             66 &            7.08  &         44.22 &             0.020 & -1.03\\
& Low-$\lambda_{Edd}$   &             73 &            8.16  &         44.17   &             0.022 & -2.21 \\

  \hline

& High-$L$ &             72 &            8.02 &         44.96 &             0.042  & -1.20\\
& Low-$L$ &             66 &            7.89   &         44.49 &             0.043 & -1.51\\
 $0.03<z<0.06$ & Large  M$_{\rm BH}$      &             65 &            8.35   &         44.79 &             0.041 & -1.80 \\
& Small  M$_{\rm BH}$  &             69 &            7.33  &         44.65 &             0.043 & -0.87\\
& High-$\lambda_{Edd}$ &             70 &            7.36  &         44.76&             0.041 & -0.83\\
& Low-$\lambda_{Edd}$   &             66 &            8.34  &         44.66   &             0.043 & -1.82\\

\hline
\end{tabular}
   \caption{Median parameter values of the BASS AGN subsamples for which we perform the $k$NN measurements.}
\label{tab:tab1}
\end{table*}

\subsection{Galaxy Sample}
\begin{figure}
  \centering

    \includegraphics[width=0.49\textwidth]{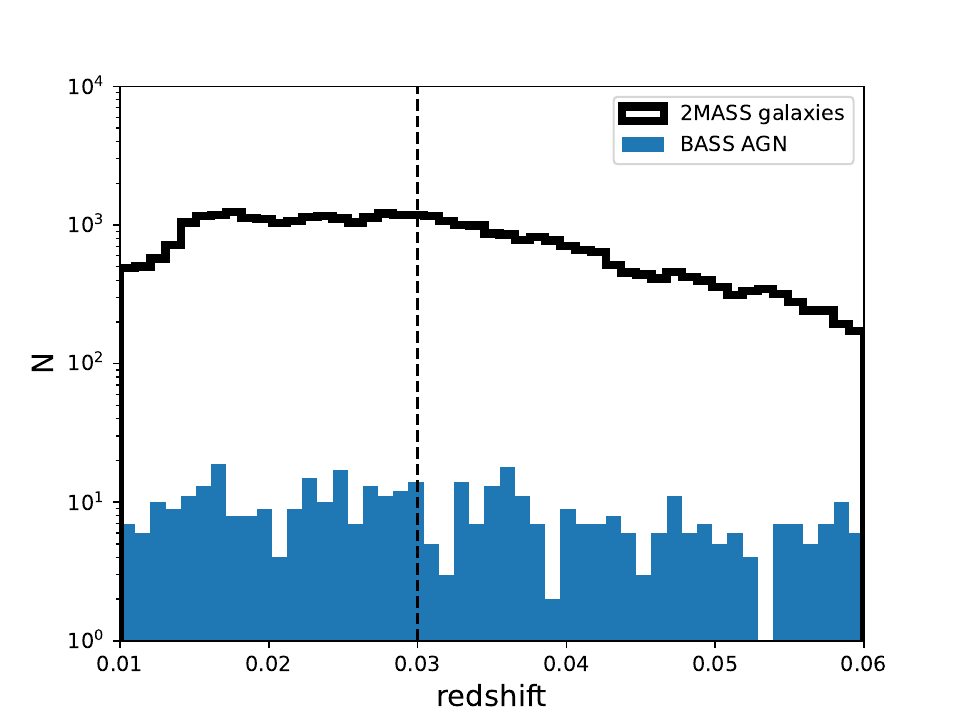}
  \caption{Redshift distributions of the full AGN sample (blue) and the galaxy catalog(black). The redshift separating our two redshift ranges (0.03) is shown by the dotted line.}
  \label{fig:AGN-gal-distribution}
\end{figure}

The tracer galaxy sample was taken from the 2MASS Redshift Survey (2MRS; \citealt{Huchra:2012}), which includes over 40k near-infrared-selected galaxies with $K_s<11.75$ and spectroscopic redshifts. They cover nearly the full sky (excluding the galactic plane ($|b|>8^{\circ}$). We selected 36,584 in the same redshift range as the AGN sample: $0.01<z<0.06$. We excluded from the catalog the counterparts of the BASS AGN (within 3") so that the first nearest-neighbors are separate galaxies and not the BASS AGN themselves. Fig. \ref{fig:AGN-gal-distribution} shows the redshift distribution of the 2MASS galaxies compared to the BASS AGN. 

Due to the shallow flux limit of 2MRS, the galaxy number density decreases by $88\%$ between our two redshift ranges. Therefore, each redshift range is sensitive to different environments and spatial scales; at lower$-z$, the survey is more complete and so galaxy neighbors trace more diverse regions of the cosmic web. The higher redshift range is limited to the brightest galaxies, so the neighbor distances are typically larger and the galaxies are less likely to probe the more typical field environments. We show the effect of the flux limit on $k$NN distances using a volume-limited AGN sample in Appendix \ref{sec:appendixA}. As a result of this flux limit, we expect less significant AGN $k$NN trends in the higher-z range due to poorer statistics and less diverse environments probed.
We emphasize that since the primary focus of this work is on the relative differences in the clustering signal between AGN subsamples within the same redshift range, this flux limit does not bias the trends with AGN properties. For each comparison, the subsamples (each controlled for redshift) are cross-correlated with the same galaxy samples. 
\\

\section{$k$th-nearest-neighbor calculation}
\label{sec:methods}

$k$-Nearest Neighbor statistics refer to cumulative probability distributions of the distances between a set of query points (either a random sample in the volume or a real galaxy sample) and their nearest galaxy neighbors. It is sensitive to all orders of the correlation function \citep{Banerjee_2020}, and it is especially powerful on small, non-linear scales. While the original approach utilizes a random sample of query points, useful for probing underdense regions and testing cosmology \citep{Banerjee_2020, Wang_2022}, the 'Data-Data' approach instead uses the galaxy sample itself as query points, therefore being more sensitive to the overdense regions of the cosmic web \citep{Yuan_2023}. For the first time, we perform a cross-correlation between X-ray AGN and galaxies to calculate the AGN-galaxy $k$NN.

Following \cite{Yuan_2023, Banerjee_2020, Banerjee_2022}, we measure the angular distances from each BASS AGN in our sample to their $k$th nearest 2MRS galaxy neighbor, where $k$ ranges from 1 to 7. By collecting these distances for each AGN in this range of $k$'s, we construct empirical cumulative distribution functions (CDFs). These CDFs correspond to the probabilities that AGN have their $k$th nearest neighbor within a given angular distance.
 The distances to the $k$-th nearest neighbor from each AGN is calculated using the \texttt{Balltree} algorithm from the \texttt{sklearn} Python package. 
A histogram of the distances using the specified bins is computed for each neighbor, aggregating via a cumulative sum to obtain the CDF.

We sample the CDFs with several values of angular separations. In this study, we chose to use six total angular bins to construct the CDFs for the full AGN sample and five total bins when splitting the sample by a given parameter. These numbers were chosen as a balance between having sufficient samplings to generate a CDF and avoiding having too many that would increase noise and lead to possible singularity in the covariance matrix. Since the CDFs are normalized such that the largest bin is always one, we exclude it from the analysis. 
For the full sample of AGN, this corresponds to having 35 degrees of freedom (five per neighbor). For the other subsamples based on mass, luminosity, or Eddington ratio, we have 28 degrees of freedom (four per neighbor). 
As discussed in Section \ref{sec:Covariance}, 
these choices were made to minimize noise and maintain numerical stability in the covariance matrix.
Previous work has also shown that the covariance matrix is most reliable for CDF probabilities between 0.1 and 0.9 \citep{Yuan_2023} and so the angular bins were chosen so that the respective probabilities roughly fell in this range. To maintain this condition across increasing neighbors, the angular samplings were linearly shifted by 0.15$^{\circ}$ for each neighbor. For the first neighbor, the bins were linearly sampled between 0.1 and 1 degree.

\subsection{Covariance Matrix estimation}
\label{sec:Covariance}
The uncertainties of the $k$NN CDFs and the covariances were calculated via jackknife resampling. The survey was partitioned into 49 sky regions, with each patch containing approximately the same number of AGN and galaxies. The choice of 49 regions was made based on a balance between having sufficiently large regions and having enough samples to approximate a normal distribution to measure the uncertainties. Furthermore, we varied the number of jackknifes, M, from $5^2$ to $9^2$ to ensure that the covariance matrix, and more importantly, the chi-squared value calculated from it, remained relatively stable and was not sensitive to the number of jackknife samples. Across various measurements, we observed that the covariance matrix remained stable for $M>49$, and we found that the chi-squared values derived from it were not significantly changed by increasing $M$ past this mark.

The covariance matrix was calculated as follows:

$$C_{i,j} = \frac{M}{M-1} \sum_{n=1}^{M} \left( x_{k,i} - \bar{x}_{i} \right) \left( x_{k,j} - \bar{x}_{j} \right)$$

\noindent where M is the number of jackknife samples (49), $x_{k,i}$ is the measurement in the $i$th angular bin-neighbor pair and $n$th jackknife sample and $\bar{x}_{i}$ is the mean across all jackknifes for the $i$th angular bin-neighbor pair. Here, $i$ and $j$ span all possible combinations of angular bins and neighbors, treating each bin-neighbor pair as a distinct data point.
Therefore, $i$ and $j$ run from $1$ to $N_{k}\times N_{\theta}$, where $N_{k}$ is the number of neighbors probed, and $N_{\theta}$ is the number of angular samplings per neighbor. The errors on $x_{i}$ are the square roots of the diagonal of this matrix: $\sigma_{i} = \sqrt{C_{i,i}}$.

To ensure the stability and reliability of our measurements, we confirmed that inverting $C$ twice returned our original covariance matrix within floating point error. 
We also calculated the condition number of each covariance matrix, ensuring it is real and of reasonable magnitude. An exceedingly large condition number would indicate significant noise and instability in the inverse matrix, leading to unreliable chi-squared measurements. 

As discussed earlier, 4 angular samplings per neighbor were chosen ($28$ degrees of freedom). This number of angular samplings is chosen to sufficiently sample the CDFs while ensuring that the number of independent elements in the covariance matrix is smaller than the number of jackknifes. Specifically, we have $28$ independent elements, sufficiently smaller than our number of jackknifes ($49$), ensuring statistical reliability.

\subsection{Correlated Chi Squared Calculation}
We aim to compare $k$NN measurements of two subsamples of AGN, binned either by $M_{\rm BH}$, $\lambda_{Edd}$ or $L_{\rm bol}$, to investigate whether or not the two subsamples are clustered consistently. To estimate the significance of the differences, we 
calculated the correlated chi-squared between $k$NN outputs for two samples using their covariance matrices, calculated as follows:

$$ \chi^2 = (\mathbf{d_1} - \mathbf{d_2})^T \mathbf{C}^{-1} (\mathbf{d_1} - \mathbf{d_2})$$

\noindent where $\mathbf{d_1}$ and $\mathbf{d_2}$ are the vectors containing the $k$NN outputs for each subsample, and $\mathbf{C}^{-1}$ is the inverse of the combined covariance matrix (the sum of the two individual matrices).

We empirically determined the null chi-squared distribution by randomly splitting the AGN sample into subsamples with the same number of AGN (68) as in the average of each defined parameter bin ($L_{\rm bol}$, M$_{BH}$, $\lambda_{Edd}$). The chi-squared values between the $k$NN statistics of the subsamples were calculated for each split. This was done 1000 times for each redshift range to characterize the distributions of chi-squared values for random splits, which we used to estimate the significance of our calculated chi-squared values. Differences between the $k$NN CDFs of examples of randomly split subsamples are shown in Appendix \ref{sec:appendixB}.

\subsection{Random AGN generation}

We calculate $k$NN statistics for the full sample of AGN compared to a randomly distributed sample, in order to show the excess probability of close galactic neighbors associated with AGN clustering. 

We follow the method of \citealt{Powell:2022} to generate randoms that consider the BASS survey sensitivity map. First, we randomly assigned RA and DEC over the sphere of the sky, excluding the galactic plane ($|b|>8^{\circ}$). Then, we assigned fluxes to each random point following the Log N-log S of the survey. We then disregarded the random values whose fluxes were below the survey sensitivities at their angular positions. Finally, redshifts were randomly assigned to the remaining randoms, drawn from the smoothed redshift distribution of the data. The random catalog was constructed to have the $\sim 10\times$ the number of objects as the AGN. The errors for the random $k$NN CDF's were calculated via the standard deviation of 25 random sample realizations.

    \begin{figure*}
      \centering

    \includegraphics[width=\textwidth]{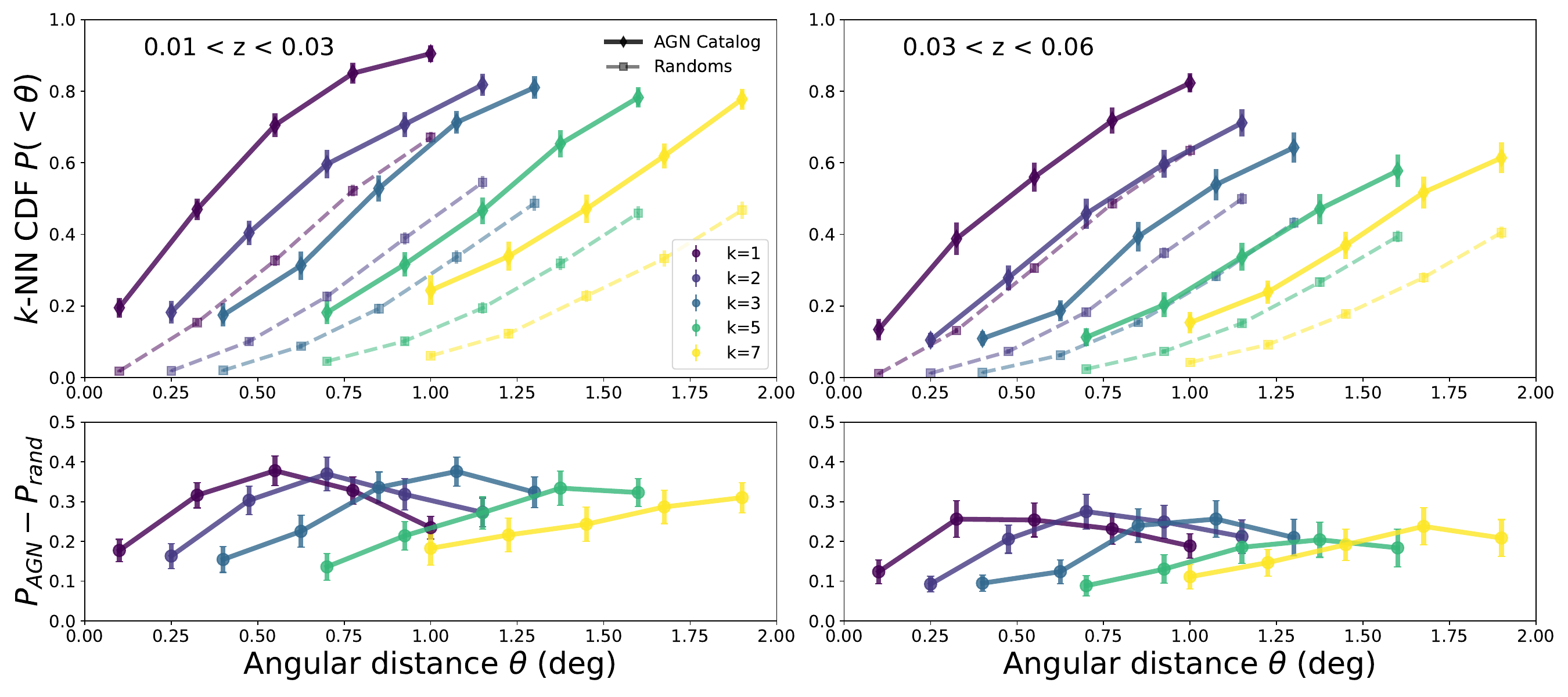}
      \caption{{\it Top panels:} Cumulative Distribution Functions (CDFs) of nearest galaxy angular separations for the AGN (solid lines) compared to randomly generated points (dashed lines) within two redshift ranges: $0.01<z<0.03$ (left) and $0.03<z<0.06$ (right). Each color corresponds to a $k^{th}$ nearest neighbor, from the 1st nearest ($k=1$; purple) to the 7th nearest ($k=7$; yellow). For clarity, the 4th and 6th neighbors are excluded. {\it Bottom panels: } The random $k$NN CDF subtracted from the AGN CDF, showing the excess probabilities for AGN to have close galactic neighbors. }
      \label{fig:agn-randoms}
    \end{figure*}

\section{BASS $k$NN Results}
\label{sec:results}
\subsection{Full AGN Sample vs. Randoms}

Figure \ref{fig:agn-randoms} presents the Cumulative Distribution Functions (CDFs) of the neighbor distances for the full AGN catalog and the randomly generated points for nearest neighbors. The AGN catalog has higher probabilities of having its galaxy neighbors within a given angular distance compared to the random sample for each $k^{th}$ neighbor. This shows that AGN are significantly more clustered than the random points, which is expected since AGNs tend to reside in more massive galaxies \citep{Powell:2018}.

Also shown in Figure \ref{fig:agn-randoms} is the difference between the AGN $k$NN CDF and that of the random catalog, which shows the excess probabilities for the neighbor separations around the AGN. The curves peak at different angular separations in both redshift ranges, showing how each neighbor is sensitive to different physical scales. We note that the absolute CDF values depend strongly on the 2MRS flux limit, the impact of which is demonstrated in Appendix \ref{sec:appendixA} for each redshift range; however, our results are concerned with the relative CDF differences in each redshift range, rather than the absolute values.

\subsection{Trends with luminosity}

    \begin{figure*}
      \centering
      \includegraphics[width=0.49\textwidth]{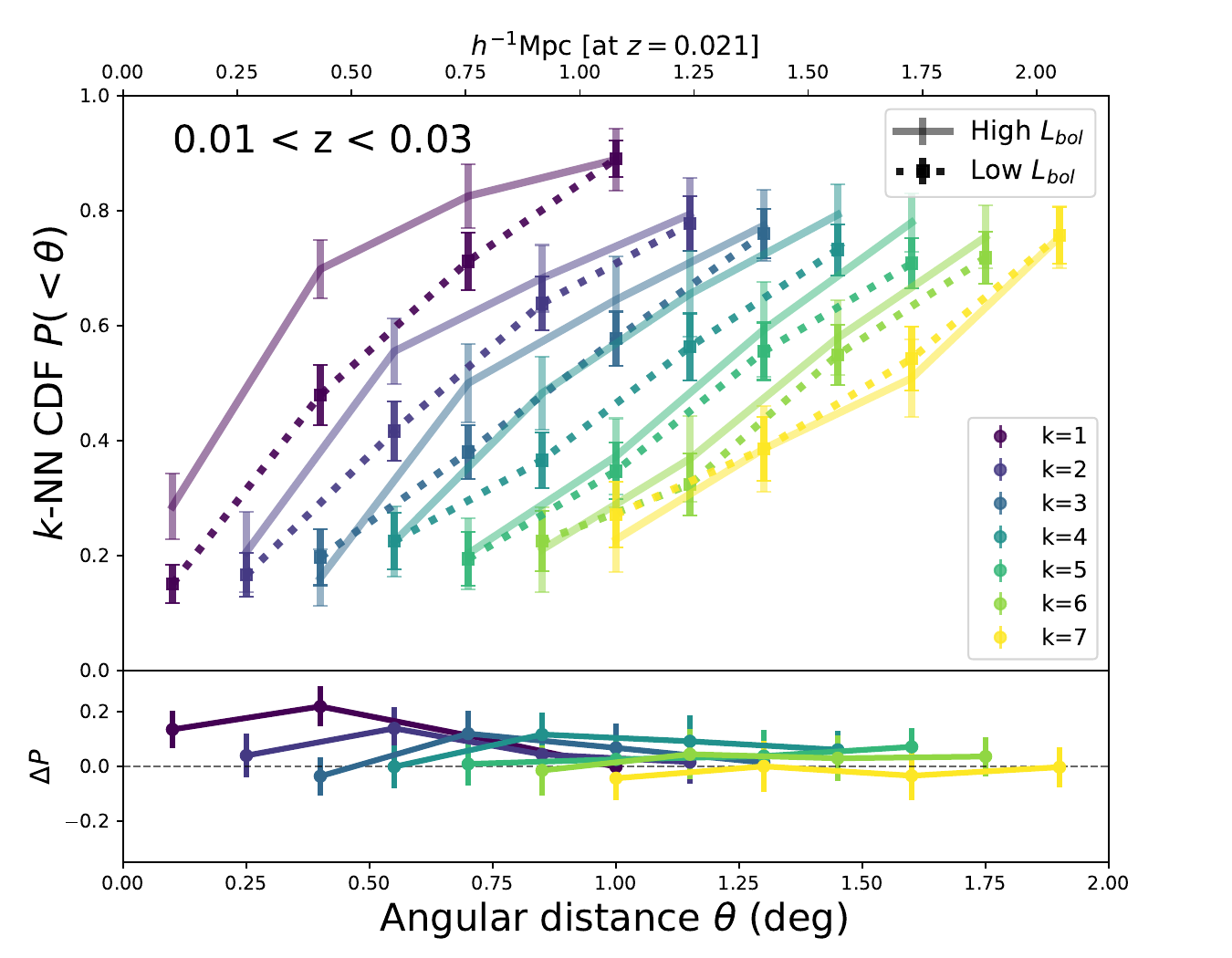}
      \includegraphics[width=0.49\textwidth]{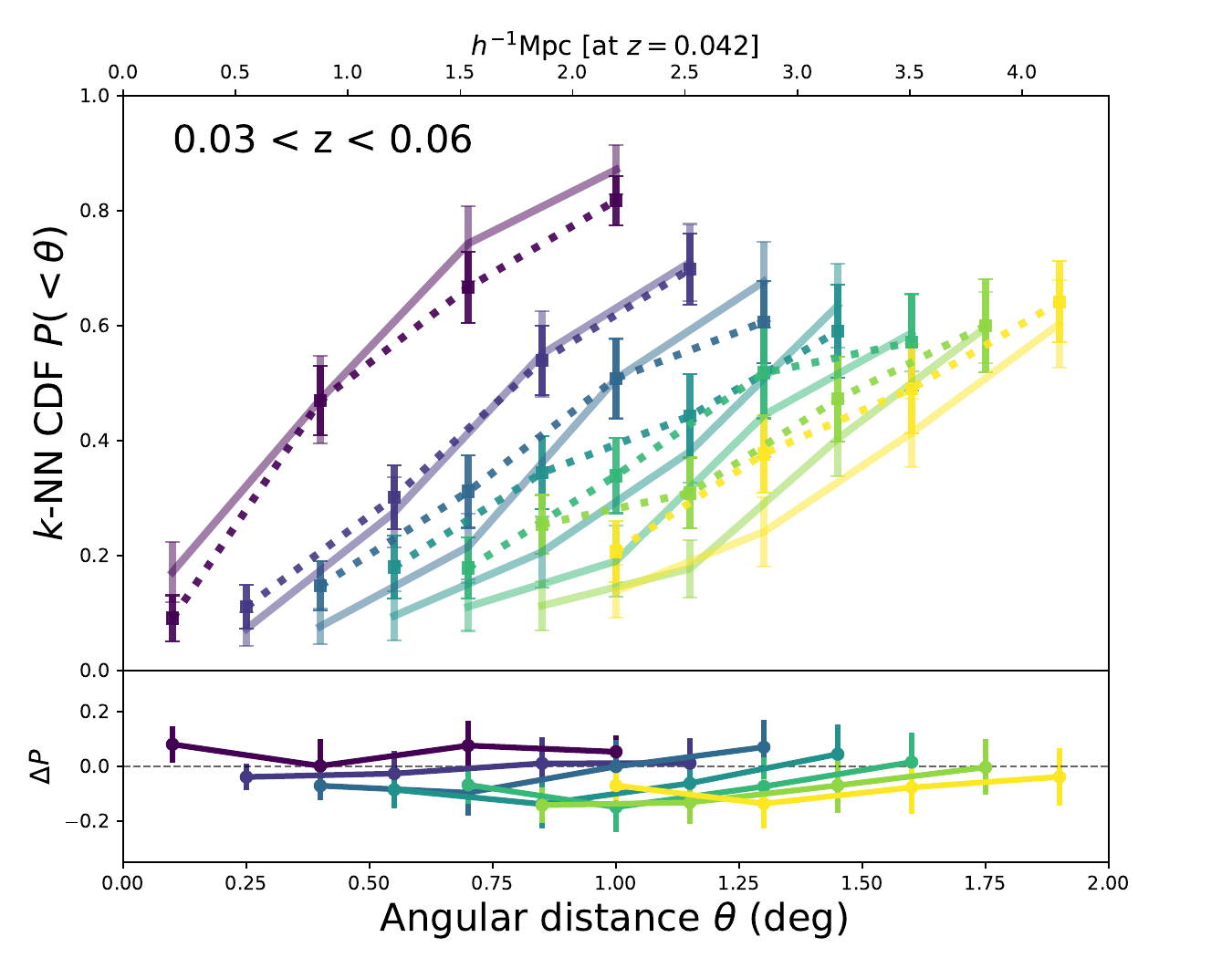}
      \caption{$k$NN CDFs of the two luminosity bins across the low redshift range (left) and higher redshift range (right). The solid lines represent the high luminosity bin and the dashed lines represent the low luminosity bin. The measurements for each galactic neighbor are marked with a different color, and the differences between the high- and low-$L_{\rm bol}$ CDFs are shown in the bottom panels. 
      }
      \label{fig:lum-CDF}
    \end{figure*}
    
We first search for trends between the AGNs extragalactic environments with their bolometric luminosities, as estimated by their hard X-ray emission. In Figure \ref{fig:lum-CDF}, we show the $k$NN Cumulative Distribution Functions for the first 7 neighbors for both luminosity bins in each redshift range, as well as the differences between the high-$L_{\rm bol}$ and low-$L_{\rm bol}$ CDFs.
We find that the higher luminosity bin has, in general, closer galactic neighbors than the low-$L_{\rm bol}$ bin, as indicated by the higher amplitudes of their CDFs. This is especially the case in the lower-$z$ range on small scales for the first and second nearest neighbor ($k=1$ and $k=2$), with the differences becoming less significant for higher $k$ neighbors.
The less-significant trends in the higher-redshift bin are likely due to the sparser galaxy sample, as well as the smaller difference in luminosity between the bright and faint subsamples within this redshift range (as shown in Fig. \ref{fig:mbh-luminosity-distribution}). It should also be noted that the lower luminosity bin in this higher-$z$ range is similar to the high-luminosity bin in the lower-$z$ sample due to the strong trend between luminosity and redshift in the BASS sample.
We calculated a correlated chi-squared of 51.21 between the subsamples for the low redshift range and 39.47 for the high redshift range (28 degrees of freedom each). 

Bolometric luminosity is a product of two fundamental AGN properties, namely black hole mass and Eddington ratio. Therefore, we next investigate trends between the AGN $k$NN statistics and those properties to determine whether or not the subtle environmental trends with luminosity are due to more significant trends with either M$_{\rm BH}$ or $\lambda_{Edd}$.

\subsection{Trends with black hole mass}
\begin{figure*}
      \centering
        \includegraphics[width=0.49\textwidth]{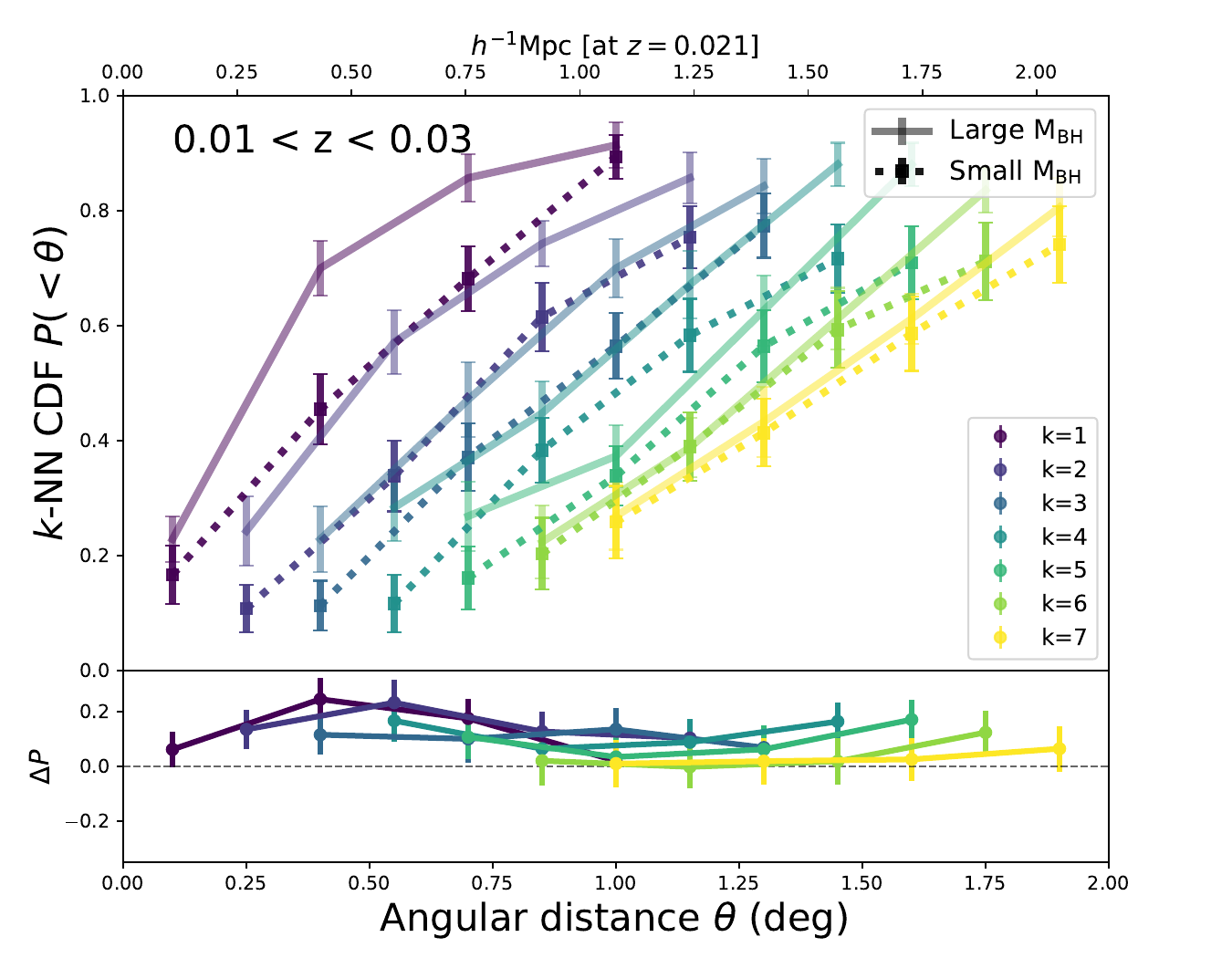}
      \includegraphics[width=0.49\textwidth]{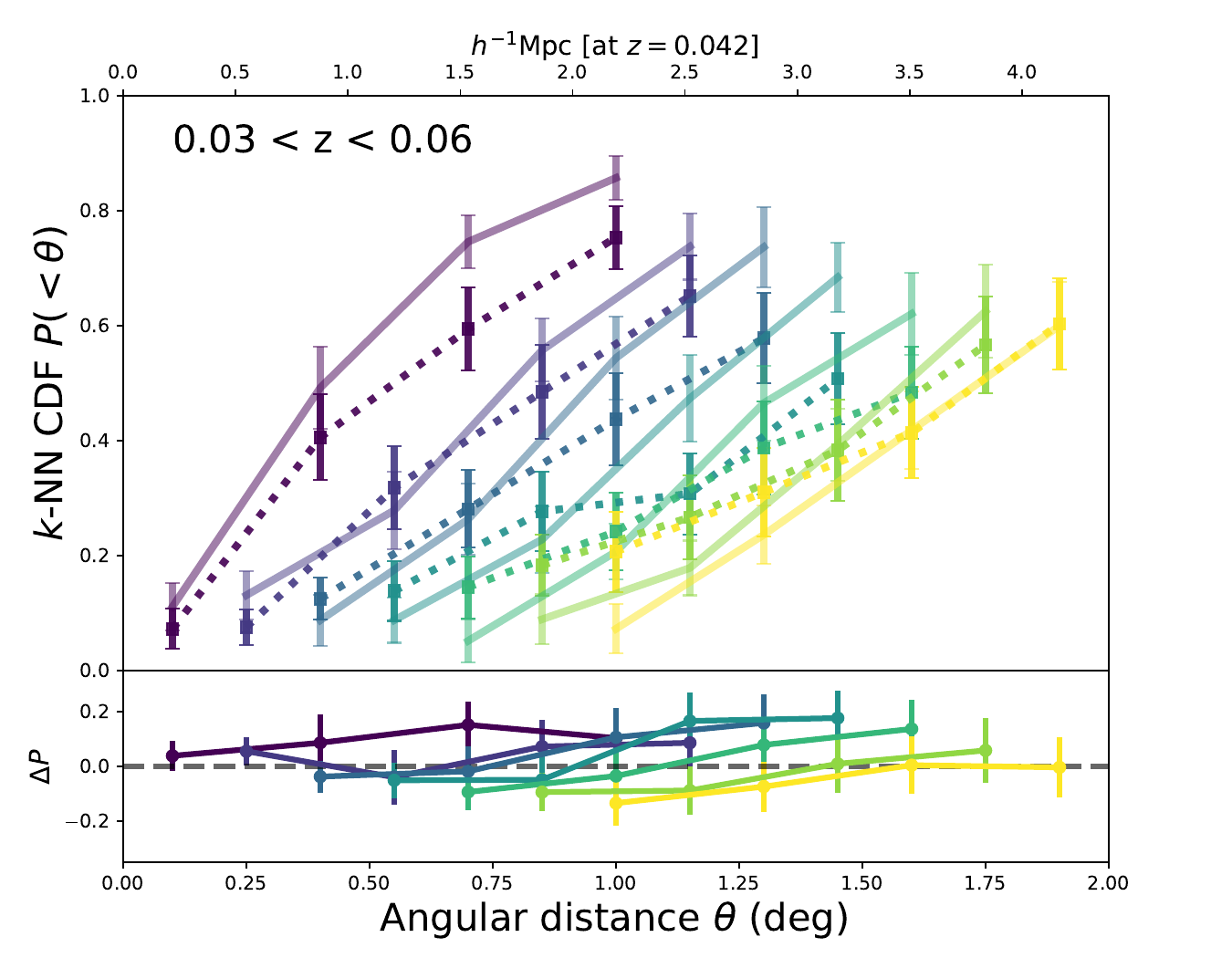}
      \caption{$k$NN CDFs of the two black hole mass bins for the low redshift range (left) and higher redshift range (right). The solid lines correspond to the high-BH mass AGN and the dashed lines represent the low-BH mass subsample. The color scheme is the same as in Fig. \ref{fig:lum-CDF}, and the differences between the high- and low-$M_{\rm BH}$ CDFs are shown in the bottom panels }
      \label{fig:bh-mass-CDF}
    \end{figure*}

Figure \ref{fig:bh-mass-CDF} presents the CDFs for our two black hole mass bins for the nearest 7 neighbors. The results are again shown for both redshift ranges. In the case of the lower redshift range ($0.01<z<0.03$), the high-mass bin has higher probabilities of having closer neighbors within a given angular distance than the lower black hole mass bin for all values of $k$. The differences appear to be most significant for first three galactic neighbors ($k=$1, 2, and 3), which corresponds to scales between 0.25 and 0.75 $h^{-1}$Mpc (assuming the median redshift of the sample). However, differences are still seen for larger $k$s that correspond to scales of $\sim 1.75$ $h^{-1}$Mpc. The $\chi^{2}$ difference between the two mass bins is 99.57 (with 28 degrees of freedom).

For the higher redshift range, AGN with more massive SMBHs are still seen to have closer nearest neighbors ($k=1$), although the differences are again less significant than at lower-$z$. As previously discussed, the weaker significance is likely due to the fact that the 2MASS galaxies become less complete at $z>0.03$.
Additionally, the scales associated with these first-neighbor angular distances at this redshift range are larger; at $\sim 1$ $h^{-1}$Mpc, no longer probing scales of the 1-halo term. Therefore, if the environmental differences between high- and low-mass SMBHs are primarily on scales of the 1-halo term (as found in \citealt{Powell:2022}), we would expect weaker trends in this higher-z range. 
For $k$=5,6, and 7, we see little or no significant differences between the two M$_{\rm BH}$ CDFs. However, the high-mass bin does seem to have closer neighbors for $k=1-4$, corresponding to scales roughly from $1-3~h^{-1}$Mpc.
The $\chi^{2}$ difference between the two mass bins for all $k$ values in this higher redshift range is 55.86.

The closer neighbors for the high-mass bin correspond to more massive SMBHs being more clustered than less-massive SMBHs. This is the case in both redshift bins and was also found using samples with consistently-measured $M_{\rm BH}$ (i.e., using only $\sigma_{*}$-derived values or using only broad line measurements), shown in Appendix \ref{sec:appendixC}. It is also consistent with what is found with the correlation function for this sample, as measured in \citealt{Powell:2022}. 
However, the $k$NN difference is more significant than what was found with the 2-point function ($3.8\sigma$ compared to $2.3\sigma$; see Section \ref{sec:sigs}), showing that $k$NN statistics have additional information that may better probe small-scale clustering than the correlation function.

\subsection{Trends with Eddington Ratio}

\begin{figure*}
      \centering
      \includegraphics[width=0.49\textwidth]{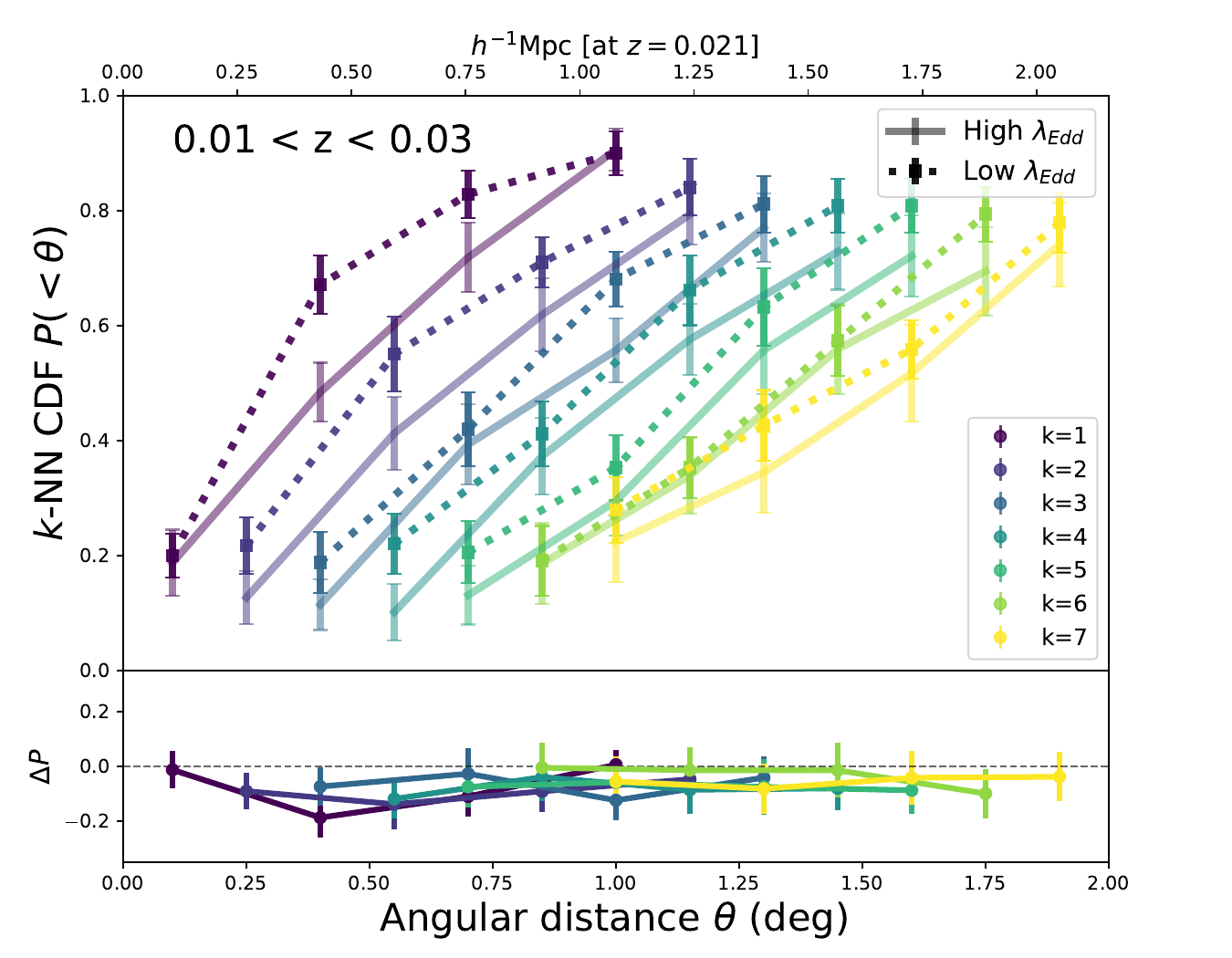}
      \includegraphics[width=0.49\textwidth]{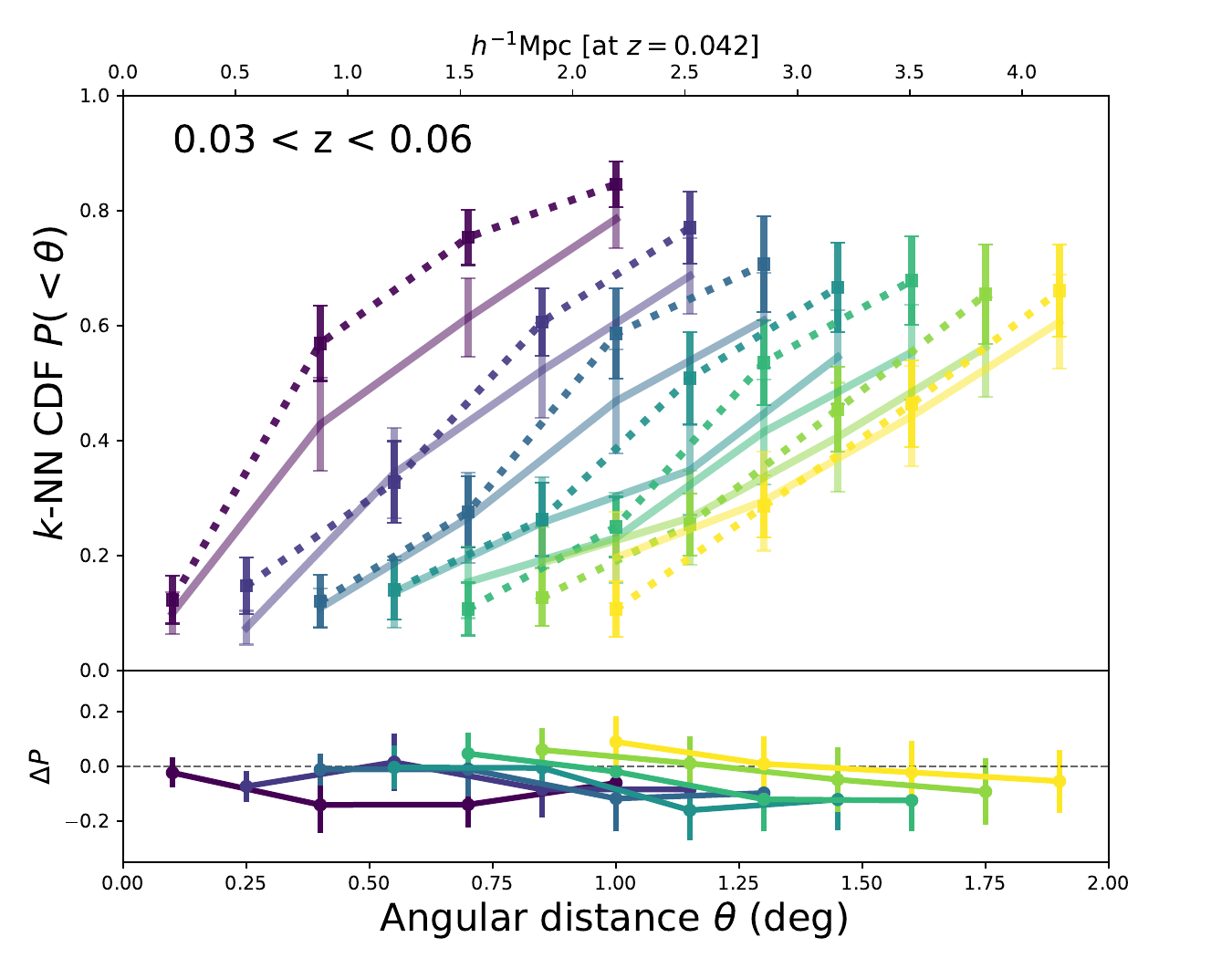}
      \caption{$k$NN CDFs of the two Eddington ratio bins for the low redshift range (left) and higher redshift range (right). The solid lines correspond to the high-$\lambda_{Edd}$ AGN and the dashed lines represent the low-$\lambda_{Edd}$ subsample. The color scheme is the same as in Figs. \ref{fig:lum-CDF} and \ref{fig:bh-mass-CDF}, and the differences between the high- and low-$\lambda_{\rm Edd}$ CDFs are shown in the bottom panels}
      \label{fig:edd-CDF}
    \end{figure*}

Finally, we show the $k$NN CDFs for the two bins of Eddington ratio in Fig. \ref{fig:edd-CDF}. For both redshift ranges, the trends are opposite to what was seen for the BH mass bins; the low-Eddington AGN have slightly closer neighbors than the high-Eddington AGN, especially for lower values of $k$. However, the differences between the $\lambda_{Edd}$ bins are less significant than when binning by BH mass; we calculate $\chi^{2}=50.82$ for the lower$-z$ range, and  $\chi^{2}=35.05$ for the higher$-z$ range. The differences lie mostly within the estimated uncertainties. We note that the $k$NN differences are not very sensitive to the bin sizes; defining the bins as being more widely separated from each other (such that there is less contamination in each due to the larger uncertainties on $\lambda_{\rm Edd}$) results in the same conclusions.

Because the trends are opposite of those found for $L_{\rm bol}$, and since $\lambda_{Edd}\approx \frac{L_{bol}}{M_{BH}}$, the differences in the $\lambda_{Edd}$ bins are most likely driven by black hole mass rather than accretion rate. Additionally, these findings are further evidence that the trends in luminosity were driven by black hole mass and not by Eddington ratio, since the differences in $\lambda_{Edd}$-binned $k$NN CDFs go in the opposite direction. We discuss this further in Section \ref{sec:discussion}.\\

\subsection{Trend significances}
\label{sec:sigs}

\begin{figure}
    \centering
    \includegraphics[width=\linewidth]{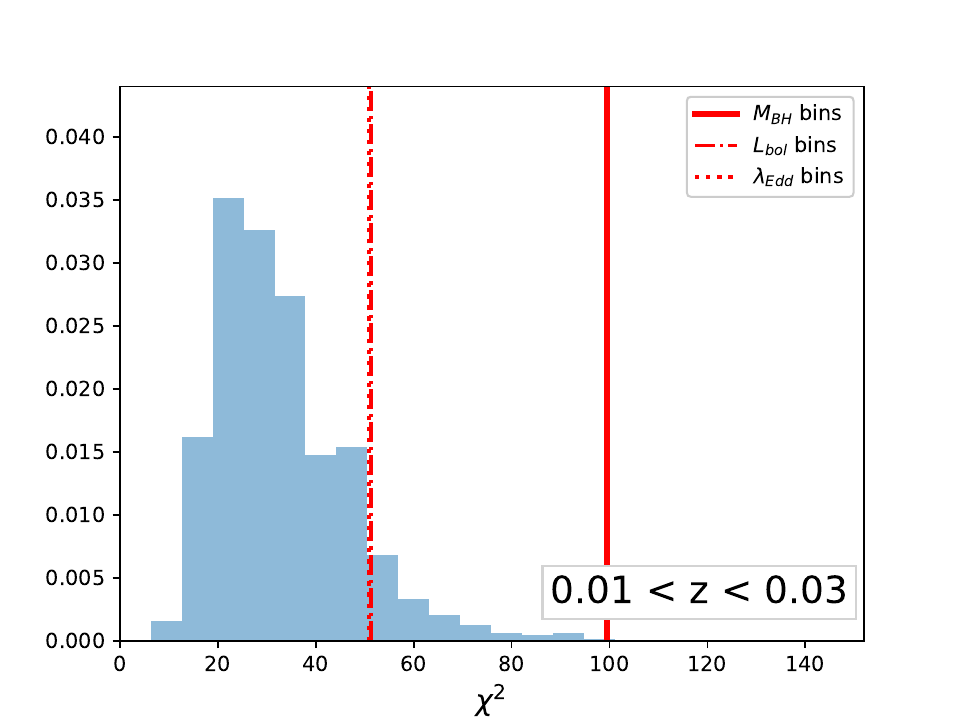}
    \includegraphics[width=\linewidth]{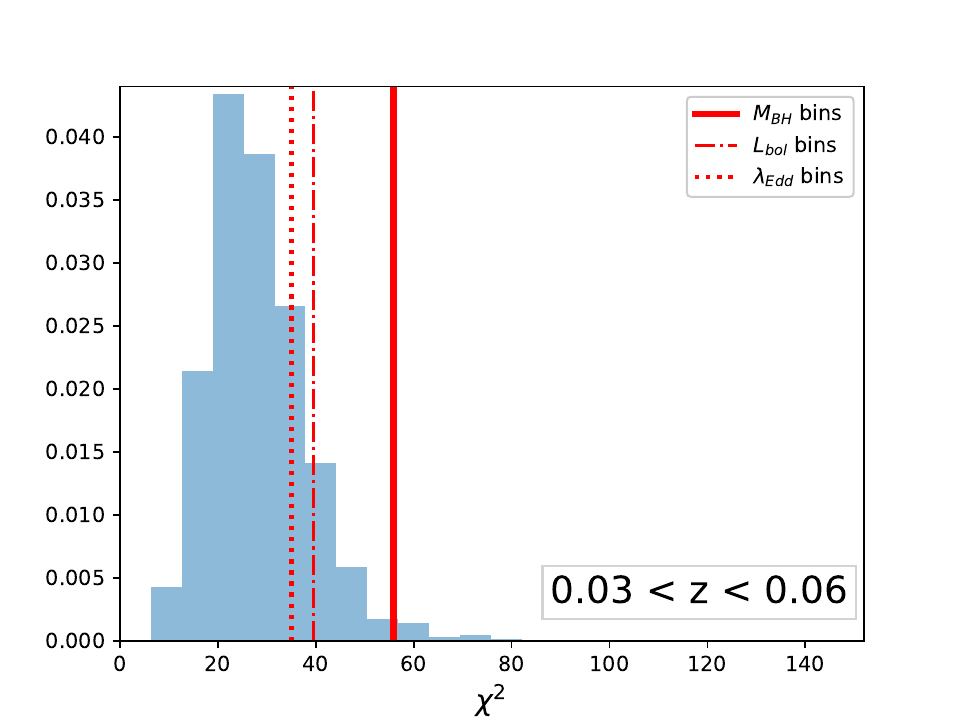}
    \caption{Comparison of our chi-squared values for each binning configuration (red lines) with the null distribution (blue histograms). The latter was obtained by randomly splitting the AGN sample into two subsamples the size of each parameter bin and comparing them. The results for our lower redshift range are shown on top, and the higher$-z$ on the bottom. }
    \label{fig:chis}
\end{figure}

The comparison of the $\chi^{2}$ values for each parameter binning with the empirically-measured null $\chi^{2}$ distribution is shown in Figure \ref{fig:chis} for both redshift ranges. The chi-squared of the lower-z $M_{\rm BH}$ bins is in the 99.98th percentile of random splits, corresponding to a $3.6-\sigma$ clustering difference. The luminosity and Eddington ratio bins, on the other hand, are in the 91st percentile. For the higher-$z$ range, the mass binning chi-squared is in the 94th percentile of the higher-$z$ null distribution, while it is 74th and 64th for the luminosity and Eddington ratio bins, respectively. Combining the two redshift ranges results in $p-$values of $8.06\times 10^{-5}$ for the mass trends, and $p-$values of 0.05 and 0.08 for the luminosity and Eddington ratio trends, respectively.

To summarize, while there are $<2\sigma$ differences in the nearest neighbor statistics of AGN binned by luminosity and Eddington ratio, $k$NN differences are much more significant when binning by black hole mass.
This may indicate that the extragalactic environment was important for earlier supermassive black hole growth, while being less important for the instantaneous accretion rate.

\begin{figure}
    \centering
    \includegraphics[width=\linewidth]{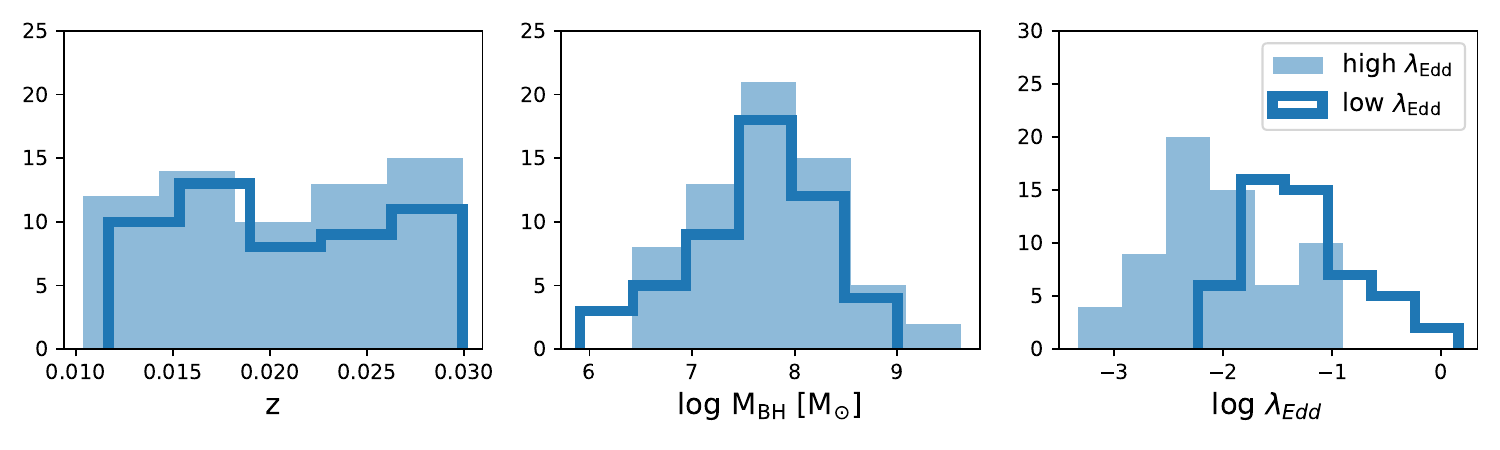}
    \includegraphics[width=\linewidth]{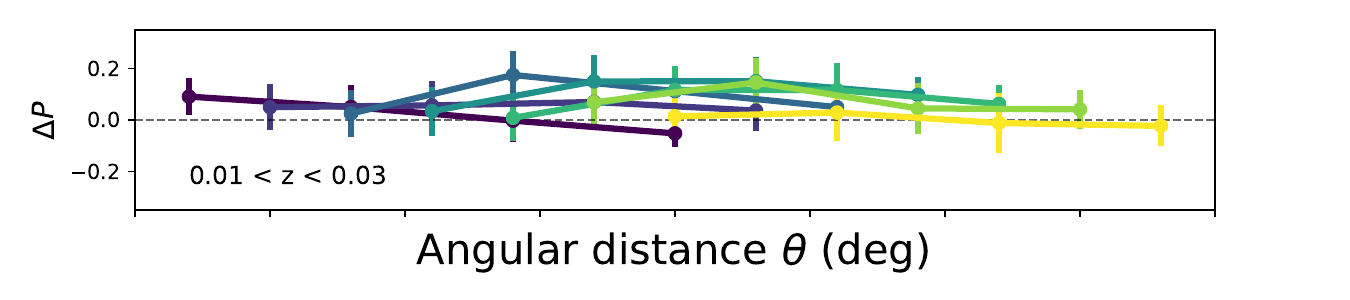}
    \caption{$k$NN statistics vs. Eddington ratio, controlled for redshift and black hole mass. Top: distributions of redshift, $M_{\rm BH}$, and $\lambda_{Edd}$ are shown for each bin (high-$\lambda_{Edd}$ in light blue, low-$\lambda_{Edd}$ in dark blue). Bottom: $k$NN CDF differences between the high-$\lambda_{Edd}$ and low-$\lambda_{Edd}$ bin for the first 7 neighbors.}
    \label{fig:$k$NNvsEdd_mbhcont}
\end{figure}

\begin{figure}
    \centering
    \includegraphics[width=\linewidth]{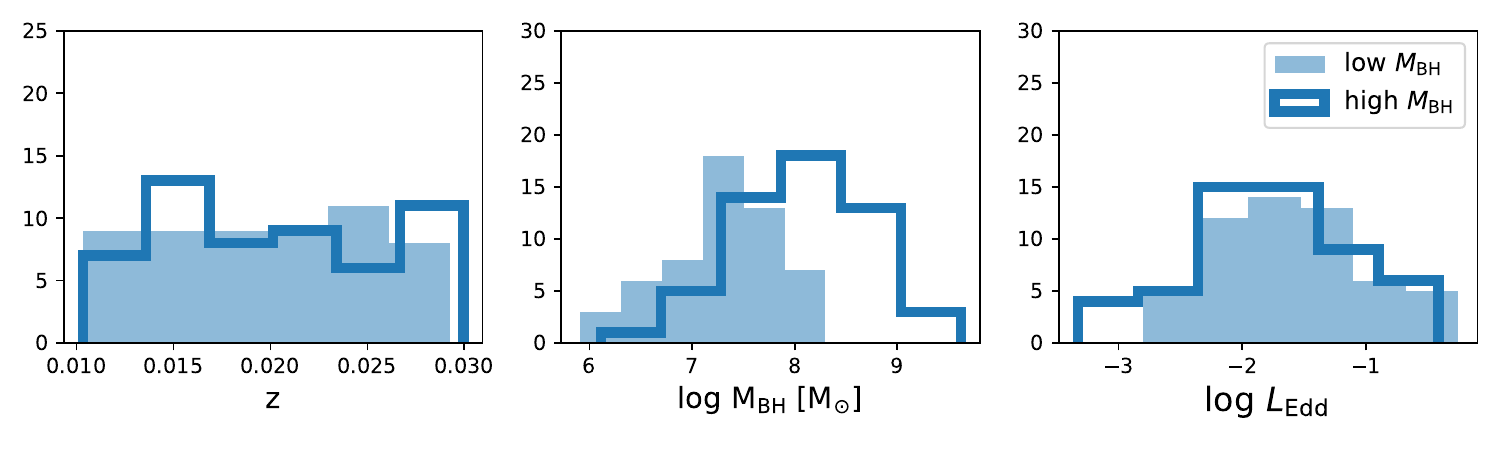}
    \includegraphics[width=\linewidth]{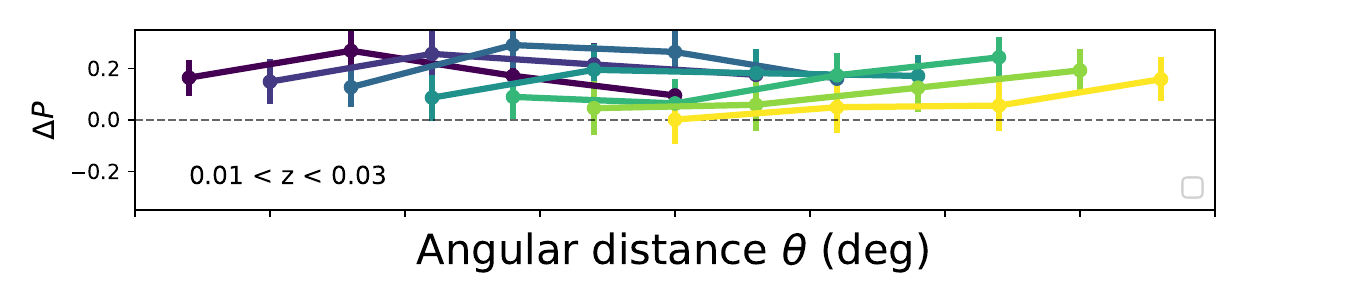}
    \caption{$k$NN statistics vs. black hole mass, controlled for redshift and Eddington ratio. Top: distributions of redshift, $M_{\rm BH}$, and $\lambda_{Edd}$ are shown for each bin (high-$M_{\rm BH}$ in light blue, low-$M_{\rm BH}$ in dark blue). Bottom: $k$NN CDF differences between the high-$M_{\rm BH}$ and low-$M_{\rm BH}$ bin for the first 7 neighbors.}
    \label{fig:$k$NNvsmbh_eddcont}
\end{figure}

To further investigate these trends, we looked at the $k$NN statistics vs. Eddington ratio while controlling for black hole mass and redshift, as well as $k$NN vs. black hole mass while controlling for Eddington ratio and redshift. We did this for the low redshift bin, where the statistics were better and where the trends were more significant. For each subsampling, we show the distributions of redshift, black hole mass, and Eddington ratio in Figures \ref{fig:$k$NNvsEdd_mbhcont} (two bins of Eddington ratio) and \ref{fig:$k$NNvsmbh_eddcont} (two bins of black hole mass). We also show the differences in the nearest-neighbor statistics in these Figures, which indicate that differences are again greater for the two $M_{\rm BH}$ bins rather than for the $\lambda_{Edd}$ bins.

\begin{figure}
    \centering
    \includegraphics[width=\linewidth]{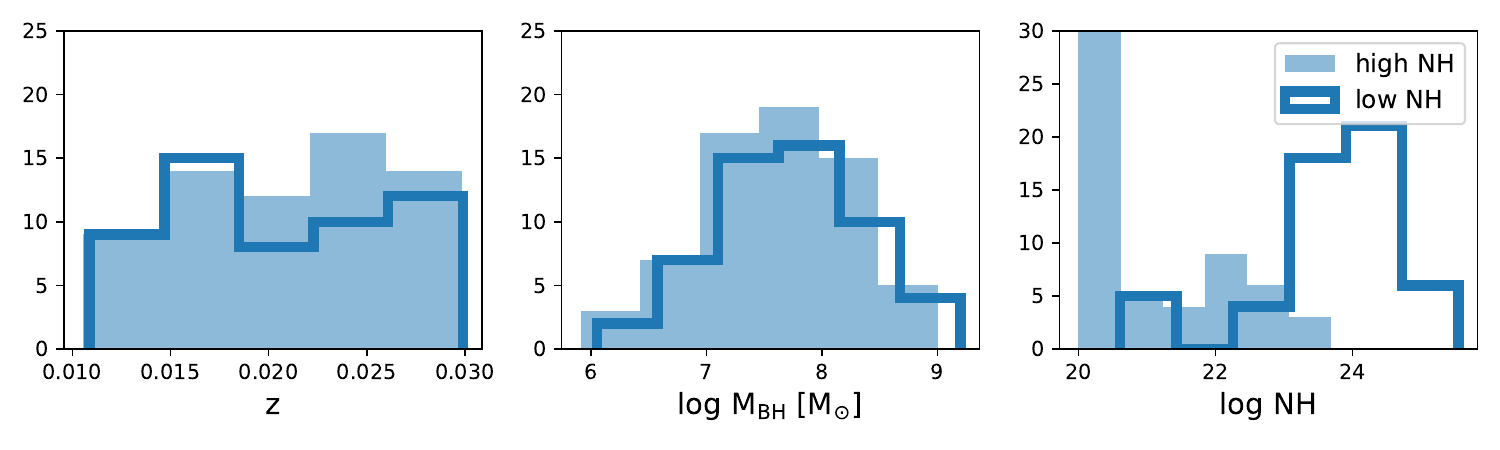}
    \includegraphics[width=\linewidth]{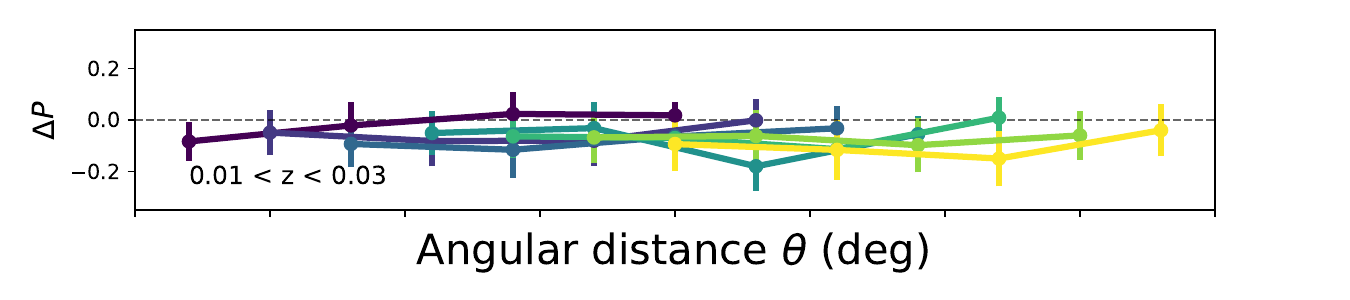}
    \caption{$k$NN statistics vs. obscuration, controlled for redshift and black hole mass. Top: distributions of redshift, $M_{\rm BH}$, and $N_{\rm H}$ are shown for each bin (high-$N_{\rm H}$ in light blue, low-$N_{\rm H}$ in dark blue). Bottom: $k$NN CDF differences between the high-$N_{\rm H}$ and low-$N_{\rm H}$ bin for the first 7 neighbors.}
    \label{fig:$k$NNvsobsc_mbhcont}
\end{figure}

Additionally, since previous results have found that AGN clustering depends on obscuring column density \citep{Krumpe:2017,Powell:2018},
we also examined possible $k$NN trends with $N_{\rm H}$ while controlling for redshift and black hole mass. We found minimal differences in the $k$NN statistics as seen in Fig \ref{fig:$k$NNvsobsc_mbhcont}. The independence of clustering with respect to obscuration indicates that observed black hole mass trends are not biased by trends with obscuration. 

To test whether these findings are the result of the well-known tight correlation between halo mass ($M_{h}$) and stellar mass (which $M_{\rm BH}$ is correlated with) or due to an additional $M_{BH}-M_{h}$ connection, we compare our results to toy models implemented into an N-body simulation, which we detail in the following section. 

\section{Interpreting $k$NN $M_{\rm BH}$ trends via N-body mocks}
\label{sec:mods}

By measuring the nearest-neighbor statistics of AGN in bins of luminosity, black hole mass, and accretion rate, we found that AGN clustering depends most strongly on $M_{\rm BH}$, while trends with Eddington ratio were less significant. This was especially true in our low-redshift range, and it indicates that environment is more related to the cumulative mass growth of the black hole rather than to the instantaneous accretion rate, the latter of which being more influenced by internal processes.
To interpret the $k$NN dependence on $M_{\rm BH}$, we compare the measurements to forward-modeled $k$NN statistics from N-body simulations, in which standard scaling relations are assumed between SMBH mass, galaxy stellar mass, and (sub)halo virial mass.

It has been well-established that SMBH mass correlates with stellar mass \citep[e.g.,][]{Kormendy:2013,Reines:2015,Shankar:2016}. While the local relation's slope, normalization, and scatter are not precisely known due to measurement uncertainties and selection biases, they have been constrained using various samples and techniques.
At the same time, the local stellar mass-(sub) halo mass relation (SMHMR) has also been constrained via abundance matching, and the scatter on that relation has also been found to be relatively small at $\sim 0.2$ dex \citep[e.g.,][]{Behroozi:2010,Moster:2013}. Since more massive galaxies and halos cluster more strongly and have closer neighbors, we test whether the $M_{\rm BH}$ $k$NN trends are due to these already-established relations, or whether additional SMBH-halo correlations are required. Evidence for correlations between SMBH mass and halo mass has been suggested in several previous works (\citealt{Marasco:2021,Powell:2022,shankar:2025}). 

We follow the method in \cite{Powell:2022}
 and \cite{Powell:2024} to populate mock black holes and galaxies within halo catalogs from a snapshot N-body simulation. For this analysis, we used the Unit simulation \citep{Chuang:2019}, which is a 1 $h^{-1}$Gpc volume box with a $1.2\times 10^{9}$ particle mass that assumed Planck cosmology \citep{Planck:2015}. We used the halo catalog from the $a=0.97810$ snapshot and the Rockstar halo finder \citep{Behroozi:2013}. 

As in \cite{Powell:2022}, we compared two models: `Model 1', in which the SMBH mass is assigned purely based on galaxy stellar mass ($M_{*}$), following a power-law mean relation with log-normal scatter; and `Model 2', which includes the same $M_{\rm BH}$-$M_{*}$ relation as Model 1, but additionally incorporates a correlation between $M_{\rm BH}$ and peak host halo virial mass ($M_{vir}$) at fixed $M_{*}$. In the latter case, $M_{\rm BH}$ is tightly correlated with halo mass. 

\subsection{Mock  generation}

Here we summarize the steps to populate the halo catalog with mock SMBHs and galaxies (additional details can be found in \citealt{Powell:2022,Powell:2024}). We also describe our method of selecting the mock data for the $k$NN calculation. 

\begin{figure}
    \centering
    \includegraphics[width=\linewidth]{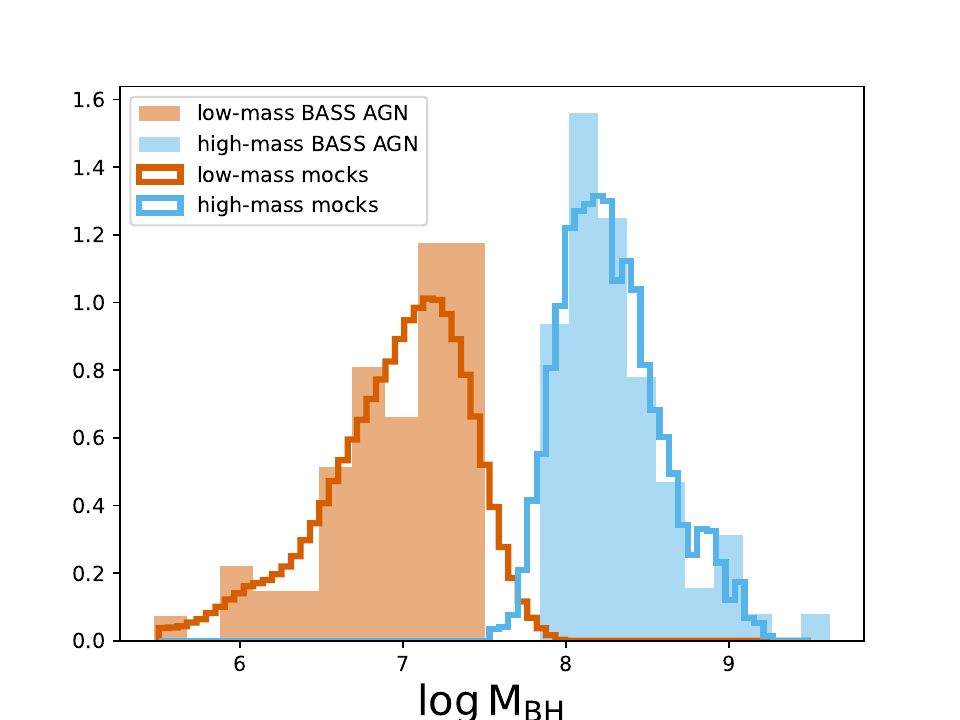}
    \caption{Normalized distributions of SMBH mass for each $M_{\rm BH}$ bin (orange and blue for the small and large bin, respectively) for the data (filled histograms) and mocks (step histograms). The mocks were chosen to match the smoothed BASS distributions.}
    \label{fig:mockmbh}
\end{figure}

\begin{enumerate}
    \item Each halo and subhalo\footnote{A subhalo is a dark matter halo that resides within the virial radius of a larger halo.} with a virial mass $M_{Vir}>5\times 10^{10}h^{-1}$M$_{\odot}$ in the halo catalog was assumed to have a mock galaxy at its center. We assigned the stellar masses based on the SMHMR from \cite{Behroozi:2010}, which includes a log-normal scatter of 0.2 dex on the mean relation. 

    \item We populated each mock galaxy with a mock SMBH with a mass assigned according to either Model 1 or Model 2. In both cases, a powerlaw relation between $M_{\rm BH}$ and $M_{\rm *}$ was assumed in the form $\log M_{\rm BH}=$ norm $+$ slope $\times ~ \log_{10}(M_{*}/10^{11}M_{\odot})$, with a uniform lognormal scatter. The results were compared for three different sets of parameters for this relation. In the case of Model 2, we use the \texttt{ conditional}\_{\tt abunmatch} function from {\tt halotools} to assign $M_{\rm BH}$ such that, for fixed stellar mass, there is a 1-1 relation between $M_{\rm BH}$ and peak virial (sub)halo mass, resulting in a tighter overall relationship between black hole mass and halo mass. 

    \item We then transformed coordinates such that the observer was at the center of the simulation box. Using the 3D positions and line-of-sight velocities of the host (sub)halos, we selected the mock SMBHs with `redshifts' between 0.01 and 0.03 to match the data. We then downsampled these mocks to create two black hole mass bins, each with the same distribution of $M_{\rm BH}$ as the data bins. To achieve this, we divided the smoothed data $M_{\rm BH}$ histogram with that of the mocks, and then normalized and interpolated it to define the completeness curve ($f$) as a function of $M_{\rm BH}$. We then randomly assigned each mock SMBH a number between 0 and 1; if the number was below the $f$ value for its mass, we kept the mock in the sample; otherwise, we masked it. This was done for both large and small $M_{\rm BH}$ bins. The resulting mass distributions of the mocks and data are shown in Fig. \ref{fig:mockmbh} for both mass bins.
\end{enumerate}

For each model, we assumed three different parameterizations for the $M_{\rm BH}-M_{*}$ relation based on results from the literature: \cite{Reines:2015} (norm $= 7.45$, slope $=1.05$, scatter $=0.3$), \cite{Shankar:2016} (norm $= 7.574$, slope $=1.946$, scatter $=0.32$), and \cite{Powell:2022} (norm $= 7.91$, slope $=0.93$, scatter $=0.35$).

The mock $k$NN calculation requires a mock sample of galaxies with the same number density and environments as the 2MASS galaxies. To make the mock galaxy sample, we again followed the method in \cite{Powell:2022}, in which mock galaxies were assigned $K-$band luminosities based on abundance matching using the \cite{Jones:2006} $K$-band luminosity function. This method has been shown to be able to reproduce the autocorrelation function of the 2MRS galaxies \citep{Powell:2018,Powell:2022}. However, instead of downsampling this mock catalog to match the total luminosity distribution of the data, as was done for the 2-point correlation function measurement, we broke the mocks into 6 redshift bins from $0.01<z<0.03$ (again defining the center of the simulation box as the observer). We then matched the luminosity distribution of the data {\it in each $z$-bin}. This was done to better simulate the flux-limited 2MASS survey, where the luminosity and number density strongly depend on redshift.
We verified that the number of mock galaxies in each redshift bin equaled that of the data.

 \subsection{Mock $k$NN calculation and results}

With the mock AGN and mock galaxy sample, we used the same method described in Section \ref{sec:methods} to calculate the CDFs for the first 7 neighbors. We also used the same angular bins as previously described.

\begin{figure*}
     \centering
     \includegraphics[width=\textwidth]{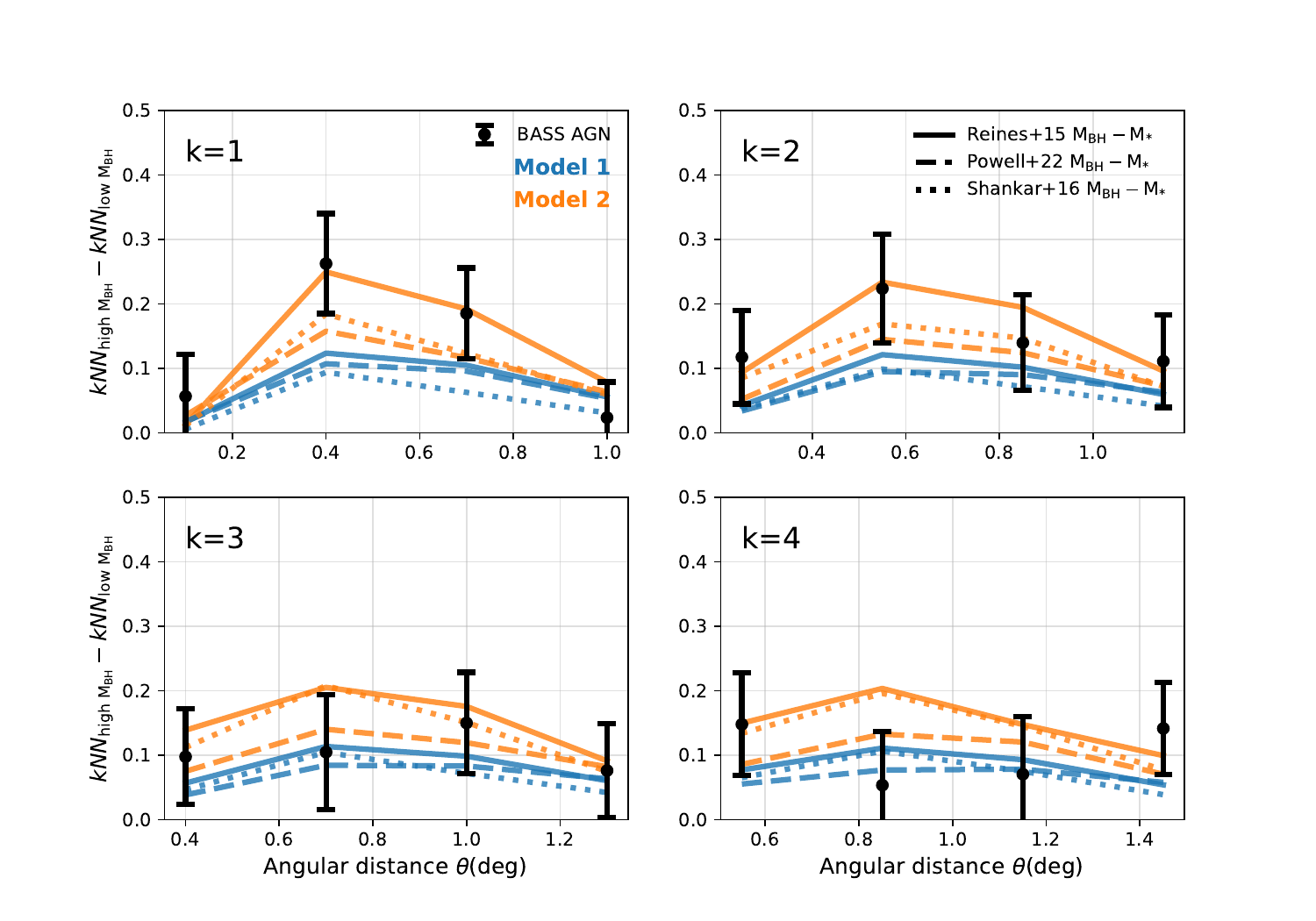}
     \caption{Differences between the high-mass and low-mass CDFs for the first 4 neighbors. The BASS measurements (black data points) are compared to two toy models: Model 1 (blue), which assumes standard $M_{\rm BH}-M_{*}$ and $M_{*}-M_{\rm halo}$ relations, and Model 2 (orange), which additionally includes a $M_{\rm BH}-M_{\rm halo}$ correlation at fixed $M_{*}$. Different line styles correspond to different assumptions for the $M_{\rm BH}-M_{*}$ relation. } 
     \label{fig:mock$k$NN}
 \end{figure*}

The differences between the high- and low-mass CDFs for each model and for the different $M_{\rm BH}-M_{*}$ parameterizations are plotted in Figure \ref{fig:mock$k$NN} for the first 4 neighbors since those had the largest differences. As expected, the differences between the two $M_{\rm BH}$ bins are higher for Model 2, in which black hole mass is strongly correlated with halo mass. However, the trends also depend on the specific $M_{\rm BH}-M_{*}$ relation assumed. The trends with $M_{\rm BH}$ are stronger for the \cite{Reines:2015} parameterization, which assumes a slightly smaller scatter (0.3 dex) than the other two (0.32 and 0.35 dex).

We find that for the first two neighbors, Model 2 better reproduces the angular distance differences than Model 1. However, this is not necessarily the case for further neighbors (i.e., higher values of $k$). This is consistent with results from the correlation function, where Model 2 better fit the AGN clustering for each BH mass bin, but only on scales of the 1-halo term ($<1$ $h^{-1}$Mpc; \citealt{Powell:2022}).

To summarize, the differences in the nearest-neighbor statistics between large and small-mass SMBHs seem to go beyond just correlations with stellar mass on small scales. However, while the toy model that assumed a $M_{\rm BH}-M_{\rm halo}$ correlation at fixed stellar mass reproduced those differences well when implemented in N-body simulations, it did worse on larger scales (corresponding to higher $k$ values). This may indicate that more sophisticated models are required to reproduce the full clustering trends. Alternatively, secondary parameters beyond mass may drive the AGN clustering statistics. 

\section{Discussion}
\label{sec:discussion}

In this paper, we calculated the $k$NN statistics of the local BASS DR2 sample, performing a cross-correlation between AGN and 2MRS galaxies by calculating angular distances between the AGN and their first 7 galactic neighbors. We found significant differences between the neighbor statistics for two redshift-controlled bins of black hole mass, especially in our lower redshift range (0.01-0.03). The AGN with massive SMBHs were found to have closer galactic neighbors, with the differences strongest on angular scales that correspond to the 1-halo term. These findings align with previous results obtained using the two-point correlation function that used a similar AGN sample \citep{Powell:2022} as well as other samples \citep{Krumpe:2015,Krumpe:2023}, which find that X-ray selected AGN with massive SMBHs are more clustered than less-massive SMBHs.

Additionally, we found less significant trends between $k$NN statistics and the Eddington ratio or luminosity. While the more luminous AGN tended to have closer galactic neighbors, we showed that this is likely due to the $M_{\rm BH}$ dependence rather than any trends between environment and accretion rate.  
This is in agreement with previous work of X-ray AGN at higher redshift that found no relation between AGN activity and environment based on the distances to their 5th, 10th, or 20th neighbors \citep{Silverman:2009}, or based on the smoothed overdensity of galaxies \citep{Yang:2018}. Optically-identified AGN in dwarf galaxies also showed no trends between AGN fraction and environment calculated from the 5th nearest-neighbor distances \citep{Siudek:2023}.
These findings also agree with previous AGN clustering studies that have found little-to-no trends between the amplitude of the AGN correlation function and luminosity and/or Eddington ratio \citep{Allevato:2011,Krumpe:2012, Krumpe:2015,Powell:2020}.

Hydrodynamic simulations have predicted AGN activity (and therefore Eddington ratio) should be enhanced in dense environments due to galaxy interactions, as indicated by the distances to the nearest neighbors \citep{Bhowmick:2020,Kristensen:2021,Singh:2023}. Indeed, an excess of mergers and dual AGN have been found in the BASS sample \citep{Koss:2010,Koss:2011,Koss:2012:L22, Koss:2018:214a}. However, the AGN triggered by interactions represent a minority of AGN in the simulations and observations, and so the predicted clustering excesses may be smaller than the current level of observational statistics so far probed. Future surveys like \textit{Euclid} will be able to better elucidate the role of environment in triggering AGN, and how it changes over cosmic time.

Some previous studies have found hints that X-ray obscured AGN may be more clustered than unobscured AGN \citep[e.g.,][]{Krumpe:2017,Powell:2018,Koutoulidis:2018}, while others have shown they cluster consistently \citep{Viitanen:2021}. However, such trends may be partially explained by selection effects; that is, obscured AGN typically need to be intrinsically brighter in order to be detected in a given (soft) X-ray survey. Because more luminous AGN tend to host larger SMBHs, there could be a degeneracy between obscuration, luminosity, and black hole mass that contributes to the observed clustering trends found in other samples. Our analysis found no significant difference in $k$NN statistics between obscured and unobscured AGN when controlling for black hole mass and redshift. This may indicate that the previously found clustering differences between obscured and unobscured AGN are impacted (or driven) by the black hole mass dependence.

The trend that more massive SMBHs tend to have closer neighbors has implications for our understanding of SMBH growth. 
By implementing simple models into a snapshot N-body simulation that assumed empirical relationships between the masses of SMBHs, galaxies, and (sub)halos, we found that the $k$NN trends with $M_{\rm BH}$ depended both on the input $M_{\rm BH}-M_{*}$ relation, as well as the correlation between $M_{\rm BH}$ and $M_{\rm halo}$ at fixed $M_{*}$. While neither model assumed (Model 1 with no additional $M_{\rm BH}-M_{\rm halo}$ correlation, and Model 2 with a one-to-one $M_{\rm BH}-M_{\rm halo}$ relation at fixed $M_{*}$) could fully reproduce trends in the data, Model 2 was preferred on small scales (i.e., $k=1,2$) while Model 1 was preferred on larger scales ($k>4$). Better clustering statistics, as well as a better constraint on the $M_{\rm BH}-M_{*}$ scatter are required to distinguish between these models and to infer the $M_{\rm BH}-M_{\rm halo}$ connection. 

A tight relationship between SMBH mass and halo mass has been suggested in previous studies using dynamical black hole and halo mass measurements of individual nearby galaxies \citep{Ferrarese:2002,Marasco:2021}. Such a correlation between SMBH mass and their host halos may suggest that local density plays a role in SMBH growth of SMBHs during earlier epochs. Denser cosmic regions may facilitate more rapid growth of SMBH earlier in cosmic time, via the availability of more gas or more frequent galaxy interactions. 
This is consistent with some hydrodynamic simulations that show environmental AGN triggering is more enhanced at higher redshifts \citep[e.g.,][]{Bhowmick:2020}.
Similarly, it may imply that the galaxies with more massive SMBHs have undergone a recent merger that rapidly increased their SMBH mass, as those galaxies tend to have closer neighbors.
Alternatively, a direct $M_{\rm BH}-M_{\rm halo}$ correlation could suggest that AGN feedback is relevant on halo scales. It has been suggested that the total amount of kinetic feedback and halo mass together regulate the growth of the SMBH and galaxy. This picture has been motivated by several observations \citep{Chen:2020,Marasco:2021} and seen in hydrodynamic simulations as well \citep{Bower:2017,Li:2024}. Better probes of the $M_{\rm BH}-M_{\rm halo}$ relation for various galaxy types and at several cosmic epochs will further test this scenario.

\section{Summary and Conclusions}
\label{sec:summary}

In this work, we calculated the nearest-neighbor statistics of AGN from the hard-X-ray selected BASS DR2 sample using galaxies from the 2MASS redshift survey. We searched for clustering trends with luminosity, black hole mass, and accretion rate by splitting the AGN sample into several bins, and by computing angular distances to the first 7 galactic neighbors in two redshift ranges. Our main conclusions are as follows:

\begin{enumerate}
    \item Massive SMBHs have a higher probability of having  closer neighbors than less-massive SMBHs (at the $99.98\%$ confidence level). This was especially true in the lower redshift bin ($0.01<z<0.03$) and on smaller scales of the 1-halo term.
    \item Luminous AGN tend to have closer neighbors than less-luminous AGN; however, we found that these differences were primarily driven by trends with $M_{\rm BH}$ and not accretion rate. We found no significant $K$NN differences between high- and low-Eddington ratio AGN when controlling for $M_{\rm BH}$ and $z$.
    \item $M_{\rm BH}$ $k$NN differences are more significant than those found with correlation function, showing that $k$NN statistics provide additional and complementary information on the surrounding cosmic density field. 
    \item Toy models implemented into N-body simulations suggest that clustering differences with $M_{\rm BH}$ on small scales may go beyond those due to correlations with stellar mass.
\end{enumerate}

Large spectroscopic samples of AGN and galaxies will soon come from surveys like \textit{Euclid} and eROSITA/SDSS-V/4MOST. Measuring the $k$NN statistics of these samples will be able to robustly probe the SMBH-halo connection, as well as the environmental dependence of AGN activity, in several cosmic epochs. This will provide powerful constraints on AGN fueling and feedback over time.

\begin{acknowledgements}
We acknowledge the work done by the \textit {Swift}/BAT team to make this project possible. This paper is part of the \textit{Swift} BAT AGN Spectroscopic Survey. 
KO acknowledges support from the Korea Astronomy and Space Science Institute under the R\&D program (Project No. 2025-1-831-01), supervised by the Korea AeroSpace Administration, and the National Research Foundation of Korea (NRF) grant funded by the Korea government (MSIT) (RS-2025-00553982). IMC acknowledges support from ANID programme FONDECYT Postdoctorado 3230653.
\end{acknowledgements}

\bibliography{references}

\begin{thebibliography}{66}
\expandafter\ifx\csname natexlab\endcsname\relax\def\natexlab#1{#1}\fi

\bibitem[{{Aird} \& {Coil}(2021)}]{Aird:2021}
{Aird}, J. \& {Coil}, A.~L. 2021, \mnras, 502, 5962

\bibitem[{{Allevato} {et~al.}(2011){Allevato}, {Finoguenov}, {Cappelluti}, {Miyaji}, {Hasinger}, {Salvato}, {Brusa}, {Gilli}, {Zamorani}, {Shankar}, {James}, {McCracken}, {Bongiorno}, {Merloni}, {Peacock}, {Silverman}, \& {Comastri}}]{Allevato:2011}
{Allevato}, V., {Finoguenov}, A., {Cappelluti}, N., {et~al.} 2011, \apj, 736, 99

\bibitem[{{Allevato} {et~al.}(2019){Allevato}, {Viitanen}, {Finoguenov}, {Civano}, {Suh}, {Shankar}, {Bongiorno}, {Ferrara}, {Gilli}, {Miyaji}, {Marchesi}, {Cappelluti}, \& {Salvato}}]{Allevato:2019}
{Allevato}, V., {Viitanen}, A., {Finoguenov}, A., {et~al.} 2019, \aap, 632, A88

\bibitem[{{Ananna} {et~al.}(2022){Ananna}, {Weigel}, {Trakhtenbrot}, {Koss}, {Urry}, {Ricci}, {Hickox}, {Treister}, {Bauer}, {Ueda}, {Mushotzky}, {Ricci}, {Oh}, {Mej{\'\i}a-Restrepo}, {Brok}, {Stern}, {Powell}, {Caglar}, {Ichikawa}, {Wong}, {Harrison}, \& {Schawinski}}]{Ananna:2022}
{Ananna}, T.~T., {Weigel}, A.~K., {Trakhtenbrot}, B., {et~al.} 2022, \apjs, 261, 9

\bibitem[{Banerjee \& Abel(2020)}]{Banerjee_2020}
Banerjee, A. \& Abel, T. 2020, Monthly Notices of the Royal Astronomical Society, 500, 5479–5499

\bibitem[{Banerjee \& Abel(2022)}]{Banerjee_2022}
Banerjee, A. \& Abel, T. 2022, Monthly Notices of the Royal Astronomical Society, 519, 4856–4868

\bibitem[{{Baumgartner} {et~al.}(2013){Baumgartner}, {Tueller}, {Markwardt}, {Skinner}, {Barthelmy}, {Mushotzky}, {Evans}, \& {Gehrels}}]{Baumgartner:2013}
{Baumgartner}, W.~H., {Tueller}, J., {Markwardt}, C.~B., {et~al.} 2013, \apjs, 207, 19

\bibitem[{{Behroozi} {et~al.}(2019){Behroozi}, {Wechsler}, {Hearin}, \& {Conroy}}]{Behroozi:2019}
{Behroozi}, P., {Wechsler}, R.~H., {Hearin}, A.~P., \& {Conroy}, C. 2019, \mnras, 488, 3143

\bibitem[{{Behroozi} {et~al.}(2010){Behroozi}, {Conroy}, \& {Wechsler}}]{Behroozi:2010}
{Behroozi}, P.~S., {Conroy}, C., \& {Wechsler}, R.~H. 2010, \apj, 717, 379

\bibitem[{{Behroozi} {et~al.}(2013){Behroozi}, {Wechsler}, \& {Wu}}]{Behroozi:2013}
{Behroozi}, P.~S., {Wechsler}, R.~H., \& {Wu}, H.-Y. 2013, \apj, 762, 109

\bibitem[{{Bhowmick} {et~al.}(2020){Bhowmick}, {Blecha}, \& {Thomas}}]{Bhowmick:2020}
{Bhowmick}, A.~K., {Blecha}, L., \& {Thomas}, J. 2020, \apj, 904, 150

\bibitem[{{Bower} {et~al.}(2017){Bower}, {Schaye}, {Frenk}, {Theuns}, {Schaller}, {Crain}, \& {McAlpine}}]{Bower:2017}
{Bower}, R.~G., {Schaye}, J., {Frenk}, C.~S., {et~al.} 2017, \mnras, 465, 32

\bibitem[{{Caglar} {et~al.}(2020){Caglar}, {Burtscher}, {Brandl}, {Brinchmann}, {Davies}, {Hicks}, {Koss}, {Lin}, {Maciejewski}, {M{\"u}ller-S{\'a}nchez}, {Riffel}, {Riffel}, {Rosario}, {Schartmann}, {Schnorr-M{\"u}ller}, {Shimizu}, {Storchi-Bergmann}, {Veilleux}, {Orban de Xivry}, \& {Bennert}}]{Caglar:2020}
{Caglar}, T., {Burtscher}, L., {Brandl}, B., {et~al.} 2020, \aap, 634, A114

\bibitem[{{Caglar} {et~al.}(2023){Caglar}, {Koss}, {Burtscher}, {Trakhtenbrot}, {Erdim}, {Mej{\'\i}a-Restrepo}, {Ricci}, {Powell}, {Ricci}, {Mushotzky}, {Bauer}, {Ananna}, {B{\"a}r}, {Brandl}, {Brinchmann}, {Harrison}, {Ichikawa}, {Kakkad}, {Oh}, {Riffel}, {Sartori}, {Smith}, {Stern}, \& {Urry}}]{caglar:2023}
{Caglar}, T., {Koss}, M.~J., {Burtscher}, L., {et~al.} 2023, \apj, 956, 60

\bibitem[{{Cappelluti} {et~al.}(2010){Cappelluti}, {Ajello}, {Burlon}, {Krumpe}, {Miyaji}, {Bonoli}, \& {Greiner}}]{Cappelluti:2010}
{Cappelluti}, N., {Ajello}, M., {Burlon}, D., {et~al.} 2010, \apjl, 716, L209

\bibitem[{{Cappelluti} {et~al.}(2012){Cappelluti}, {Allevato}, \& {Finoguenov}}]{Cappelluti:2012}
{Cappelluti}, N., {Allevato}, V., \& {Finoguenov}, A. 2012, Advances in Astronomy, 2012, 853701

\bibitem[{{Chen} {et~al.}(2020){Chen}, {Faber}, {Koo}, {Somerville}, {Primack}, {Dekel}, {Rodr{\'\i}guez-Puebla}, {Guo}, {Barro}, {Kocevski}, {van der Wel}, {Woo}, {Bell}, {Fang}, {Ferguson}, {Giavalisco}, {Huertas-Company}, {Jiang}, {Kassin}, {Lin}, {Liu}, {Luo}, {Luo}, {Pacifici}, {Pandya}, {Salim}, {Shu}, {Tacchella}, {Terrazas}, \& {Yesuf}}]{Chen:2020}
{Chen}, Z., {Faber}, S.~M., {Koo}, D.~C., {et~al.} 2020, \apj, 897, 102

\bibitem[{{Chuang} {et~al.}(2019){Chuang}, {Yepes}, {Kitaura}, {Pellejero-Ibanez}, {Rodr{\'\i}guez-Torres}, {Feng}, {Metcalf}, {Wechsler}, {Zhao}, {To}, {Alam}, {Banerjee}, {DeRose}, {Giocoli}, {Knebe}, \& {Reyes}}]{Chuang:2019}
{Chuang}, C.-H., {Yepes}, G., {Kitaura}, F.-S., {et~al.} 2019, \mnras, 487, 48

\bibitem[{{Coil} {et~al.}(2017){Coil}, {Mendez}, {Eisenstein}, \& {Moustakas}}]{coil:2017}
{Coil}, A.~L., {Mendez}, A.~J., {Eisenstein}, D.~J., \& {Moustakas}, J. 2017, \apj, 838, 87

\bibitem[{{DeGraf} \& {Sijacki}(2017)}]{DeGraf:2017}
{DeGraf}, C. \& {Sijacki}, D. 2017, \mnras, 466, 3331

\bibitem[{{DiPompeo} {et~al.}(2017){DiPompeo}, {Hickox}, {Eftekharzadeh}, \& {Myers}}]{DiPompeo:2017}
{DiPompeo}, M.~A., {Hickox}, R.~C., {Eftekharzadeh}, S., \& {Myers}, A.~D. 2017, \mnras, 469, 4630

\bibitem[{{Ferrarese}(2002)}]{Ferrarese:2002}
{Ferrarese}, L. 2002, \apj, 578, 90

\bibitem[{{Ghosh} {et~al.}(2024){Ghosh}, {Urry}, {Powell}, {Shimakawa}, {van den Bosch}, {Nagai}, {Mitra}, \& {Connolly}}]{ghosh:2024}
{Ghosh}, A., {Urry}, C.~M., {Powell}, M.~C., {et~al.} 2024, \apj, 971, 142

\bibitem[{{Huchra} {et~al.}(2012){Huchra}, {Macri}, {Masters}, {Jarrett}, {Berlind}, {Calkins}, {Crook}, {Cutri}, {Erdo{\v g}du}, {Falco}, {George}, {Hutcheson}, {Lahav}, {Mader}, {Mink}, {Martimbeau}, {Schneider}, {Skrutskie}, {Tokarz}, \& {Westover}}]{Huchra:2012}
{Huchra}, J.~P., {Macri}, L.~M., {Masters}, K.~L., {et~al.} 2012, \apjs, 199, 26

\bibitem[{{Jones} {et~al.}(2006){Jones}, {Peterson}, {Colless}, \& {Saunders}}]{Jones:2006}
{Jones}, D.~H., {Peterson}, B.~A., {Colless}, M., \& {Saunders}, W. 2006, \mnras, 369, 25

\bibitem[{{Kormendy} \& {Ho}(2013)}]{Kormendy:2013}
{Kormendy}, J. \& {Ho}, L.~C. 2013, \araa, 51, 511

\bibitem[{Koss {et~al.}(2012)Koss, Mushotzky, Treister, Veilleux, Vasudevan, \& Trippe}]{Koss:2012:L22}
Koss, M., Mushotzky, R., Treister, E., {et~al.} 2012, \apj, 746, L22

\bibitem[{{Koss} {et~al.}(2010){Koss}, {Mushotzky}, {Veilleux}, \& {Winter}}]{Koss:2010}
{Koss}, M., {Mushotzky}, R., {Veilleux}, S., \& {Winter}, L. 2010, \apjl, 716, L125

\bibitem[{{Koss} {et~al.}(2011){Koss}, {Mushotzky}, {Veilleux}, {Winter}, {Baumgartner}, {Tueller}, {Gehrels}, \& {Valencic}}]{Koss:2011}
{Koss}, M., {Mushotzky}, R., {Veilleux}, S., {et~al.} 2011, \apj, 739, 57

\bibitem[{{Koss} {et~al.}(2017){Koss}, {Trakhtenbrot}, {Ricci}, {Lamperti}, {Oh}, {Berney}, {Schawinski}, {Balokovi{\'c}}, {Baronchelli}, {Crenshaw}, {Fischer}, {Gehrels}, {Harrison}, {Hashimoto}, {Hogg}, {Ichikawa}, {Masetti}, {Mushotzky}, {Sartori}, {Stern}, {Treister}, {Ueda}, {Veilleux}, \& {Winter}}]{Koss:2017}
{Koss}, M., {Trakhtenbrot}, B., {Ricci}, C., {et~al.} 2017, \apj, 850, 74

\bibitem[{Koss {et~al.}(2018)Koss, Blecha, Bernhard, Hung, Lu, Trakthenbrot, Treister, Weigel, Sartori, Mushotzky, Schawinski, Ricci, Veilleux, \& Sanders}]{Koss:2018:214a}
Koss, M.~J., Blecha, L., Bernhard, P., {et~al.} 2018, \nat, 563, 214

\bibitem[{{Koss} {et~al.}(2022{\natexlab{a}}){Koss}, {Ricci}, {Trakhtenbrot}, {Oh}, {den Brok}, {Mej{\'\i}a-Restrepo}, {Stern}, {Privon}, {Treister}, {Powell}, {Mushotzky}, {Bauer}, {Ananna}, {Balokovi{\'c}}, {B{\"a}r}, {Becker}, {Bessiere}, {Burtscher}, {Caglar}, {Congiu}, {Evans}, {Harrison}, {Heida}, {Ichikawa}, {Kamraj}, {Lamperti}, {Pacucci}, {Ricci}, {Riffel}, {Rojas}, {Schawinski}, {Temple}, {Urry}, {Veilleux}, \& {Williams}}]{Koss:2022b}
{Koss}, M.~J., {Ricci}, C., {Trakhtenbrot}, B., {et~al.} 2022{\natexlab{a}}, \apjs, 261, 2

\bibitem[{{Koss} {et~al.}(2022{\natexlab{b}}){Koss}, {Trakhtenbrot}, {Ricci}, {Bauer}, {Treister}, {Mushotzky}, {Urry}, {Ananna}, {Balokovi{\'c}}, {den Brok}, {Cenko}, {Harrison}, {Ichikawa}, {Lamperti}, {Lein}, {Mej{\'\i}a-Restrepo}, {Oh}, {Pacucci}, {Pfeifle}, {Powell}, {Privon}, {Ricci}, {Salvato}, {Schawinski}, {Shimizu}, {Smith}, \& {Stern}}]{Koss:2022a}
{Koss}, M.~J., {Trakhtenbrot}, B., {Ricci}, C., {et~al.} 2022{\natexlab{b}}, \apjs, 261, 1

\bibitem[{{Koutoulidis} {et~al.}(2018){Koutoulidis}, {Georgantopoulos}, {Mountrichas}, {Plionis}, {Georgakakis}, {Akylas}, \& {Rovilos}}]{Koutoulidis:2018}
{Koutoulidis}, L., {Georgantopoulos}, I., {Mountrichas}, G., {et~al.} 2018, \mnras, 481, 3063

\bibitem[{{Krishnan} {et~al.}(2020){Krishnan}, {Almaini}, {Hatch}, {Wilkinson}, {Maltby}, {Conselice}, {Kocevski}, {Suh}, \& {Wild}}]{Krishnan:2020}
{Krishnan}, C., {Almaini}, O., {Hatch}, N.~A., {et~al.} 2020, \mnras, 494, 1693

\bibitem[{{Kristensen} {et~al.}(2021){Kristensen}, {Pimbblet}, {Gibson}, {Penny}, \& {Koudmani}}]{Kristensen:2021}
{Kristensen}, M.~T., {Pimbblet}, K.~A., {Gibson}, B.~K., {Penny}, S.~J., \& {Koudmani}, S. 2021, \apj, 922, 127

\bibitem[{{Krumpe} {et~al.}(2012){Krumpe}, {Miyaji}, {Coil}, \& {Aceves}}]{Krumpe:2012}
{Krumpe}, M., {Miyaji}, T., {Coil}, A.~L., \& {Aceves}, H. 2012, \apj, 746, 1

\bibitem[{{Krumpe} {et~al.}(2018){Krumpe}, {Miyaji}, {Coil}, \& {Aceves}}]{Krumpe:2017}
{Krumpe}, M., {Miyaji}, T., {Coil}, A.~L., \& {Aceves}, H. 2018, \mnras, 474, 1773

\bibitem[{{Krumpe} {et~al.}(2023){Krumpe}, {Miyaji}, {Georgakakis}, {Schulze}, {Coil}, {Dwelly}, {Coffey}, {Comparat}, {Aceves}, {Salvato}, {Merloni}, {Maraston}, {Nandra}, {Brownstein}, {Schneider}, {SDSS-Iv Team}, \& {Spiders Team}}]{Krumpe:2023}
{Krumpe}, M., {Miyaji}, T., {Georgakakis}, A., {et~al.} 2023, \apj, 952, 109

\bibitem[{{Krumpe} {et~al.}(2015){Krumpe}, {Miyaji}, {Husemann}, {Fanidakis}, {Coil}, \& {Aceves}}]{Krumpe:2015}
{Krumpe}, M., {Miyaji}, T., {Husemann}, B., {et~al.} 2015, \apj, 815, 21

\bibitem[{{Li} {et~al.}(2024){Li}, {Chen}, {Wang}, \& {Mo}}]{Li:2024}
{Li}, H., {Chen}, Y., {Wang}, H., \& {Mo}, H. 2024, arXiv e-prints, arXiv:2409.06208

\bibitem[{{Marasco} {et~al.}(2021){Marasco}, {Cresci}, {Posti}, {Fraternali}, {Mannucci}, {Marconi}, {Belfiore}, \& {Fall}}]{Marasco:2021}
{Marasco}, A., {Cresci}, G., {Posti}, L., {et~al.} 2021, \mnras, 507, 4274

\bibitem[{{Marcotulli} {et~al.}(2022){Marcotulli}, {Ajello}, {Urry}, {Paliya}, {Koss}, {Oh}, {Madejski}, {Ueda}, {Balokovi{\'c}}, {Trakhtenbrot}, {Ricci}, {Ricci}, {Stern}, {Harrison}, {Powell}, \& {BASS Collaboration}}]{Marcotulli:2022}
{Marcotulli}, L., {Ajello}, M., {Urry}, C.~M., {et~al.} 2022, \apj, 940, 77

\bibitem[{{Mej{\'\i}a-Restrepo} {et~al.}(2022){Mej{\'\i}a-Restrepo}, {Trakhtenbrot}, {Koss}, {Oh}, {den Brok}, {Stern}, {Powell}, {Ricci}, {Caglar}, {Ricci}, {Bauer}, {Treister}, {Harrison}, {Urry}, {Ananna}, {Asmus}, {Assef}, {B{\"a}r}, {Bessiere}, {Burtscher}, {Ichikawa}, {Kakkad}, {Kamraj}, {Mushotzky}, {Privon}, {Rojas}, {Sani}, {Schawinski}, \& {Veilleux}}]{Mejia-Restrepo:2022}
{Mej{\'\i}a-Restrepo}, J.~E., {Trakhtenbrot}, B., {Koss}, M.~J., {et~al.} 2022, \apjs, 261, 5

\bibitem[{{Mendez} {et~al.}(2016){Mendez}, {Coil}, {Aird}, {Skibba}, {Diamond-Stanic}, {Moustakas}, {Blanton}, {Cool}, {Eisenstein}, {Wong}, \& {Zhu}}]{Mendez:2016}
{Mendez}, A.~J., {Coil}, A.~L., {Aird}, J., {et~al.} 2016, \apj, 821, 55

\bibitem[{{Moster} {et~al.}(2013){Moster}, {Naab}, \& {White}}]{Moster:2013}
{Moster}, B.~P., {Naab}, T., \& {White}, S. D.~M. 2013, \mnras, 428, 3121

\bibitem[{{Oogi} {et~al.}(2020){Oogi}, {Shirakata}, {Nagashima}, {Nishimichi}, {Kawaguchi}, {Okamoto}, {Ishiyama}, \& {Enoki}}]{Oogi:2020}
{Oogi}, T., {Shirakata}, H., {Nagashima}, M., {et~al.} 2020, \mnras, 497, 1

\bibitem[{{Perez} \& {Coldwell}(2022)}]{Perez:2022}
{Perez}, N.~R. \& {Coldwell}, G. 2022, \mnras, 513, 5344

\bibitem[{{Planck Collaboration} {et~al.}(2016){Planck Collaboration}, {Ade}, {Aghanim}, {Arnaud}, {Ashdown}, {Aumont}, {Baccigalupi}, {Banday}, {Barreiro}, {Bartlett}, \& et~al.}]{Planck:2015}
{Planck Collaboration}, {Ade}, P.~A.~R., {Aghanim}, N., {et~al.} 2016, \aap, 594, A13

\bibitem[{{Powell} {et~al.}(2022){Powell}, {Allen}, {Caglar}, {Cappelluti}, {Harrison}, {Irving}, {Koss}, {Mantz}, {Oh}, {Ricci}, {Shaper}, {Stern}, {Trakhtenbrot}, {Urry}, \& {Wong}}]{Powell:2022}
{Powell}, M.~C., {Allen}, S.~W., {Caglar}, T., {et~al.} 2022, \apj, 938, 77

\bibitem[{{Powell} {et~al.}(2018){Powell}, {Cappelluti}, {Urry}, {Koss}, {Finoguenov}, {Ricci}, {Trakhtenbrot}, {Allevato}, {Ajello}, {Oh}, {Schawinski}, \& {Secrest}}]{Powell:2018}
{Powell}, M.~C., {Cappelluti}, N., {Urry}, C.~M., {et~al.} 2018, \apj, 858, 110

\bibitem[{{Powell} {et~al.}(2024){Powell}, {Krumpe}, {Coil}, \& {Miyaji}}]{Powell:2024}
{Powell}, M.~C., {Krumpe}, M., {Coil}, A., \& {Miyaji}, T. 2024, \aap, 686, A57

\bibitem[{{Powell} {et~al.}(2020){Powell}, {Urry}, {Cappelluti}, {Johnson}, {LaMassa}, {Ananna}, \& {Kollmann}}]{Powell:2020}
{Powell}, M.~C., {Urry}, C.~M., {Cappelluti}, N., {et~al.} 2020, \apj, 891, 41

\bibitem[{{Reines} \& {Volonteri}(2015)}]{Reines:2015}
{Reines}, A.~E. \& {Volonteri}, M. 2015, \apj, 813, 82

\bibitem[{{Ricci} {et~al.}(2017){Ricci}, {Trakhtenbrot}, {Koss}, {Ueda}, {Del Vecchio}, {Treister}, {Schawinski}, {Paltani}, {Oh}, {Lamperti}, {Berney}, {Gandhi}, {Ichikawa}, {Bauer}, {Ho}, {Asmus}, {Beckmann}, {Soldi}, {Balokovi{\'c}}, {Gehrels}, \& {Markwardt}}]{Ricci:2017B}
{Ricci}, C., {Trakhtenbrot}, B., {Koss}, M.~J., {et~al.} 2017, \apjs, 233, 17

\bibitem[{{Ricci} {et~al.}(2015){Ricci}, {Ueda}, {Koss}, {Trakhtenbrot}, {Bauer}, \& {Gandhi}}]{Ricci:2015}
{Ricci}, C., {Ueda}, Y., {Koss}, M.~J., {et~al.} 2015, \apjl, 815, L13

\bibitem[{{Shankar} {et~al.}(2025){Shankar}, {Bernardi}, {Roberts}, {Arana-Catania}, {Grubenmann}, {Habouzit}, {Smith}, {Marsden}, {Varadarajan}, {Tetilla}, {Angl{\'e}s-Alc{\'a}zar}, {Boco}, {Farrah}, {Fu}, {Haniewicz}, {Lapi}, {Lovell}, {Menci}, {Powell}, \& {Ricci}}]{shankar:2025}
{Shankar}, F., {Bernardi}, M., {Roberts}, D., {et~al.} 2025, \mnras [\eprint[arXiv]{2505.02920}]

\bibitem[{{Shankar} {et~al.}(2016){Shankar}, {Bernardi}, {Sheth}, {Ferrarese}, {Graham}, {Savorgnan}, {Allevato}, {Marconi}, {L{\"a}sker}, \& {Lapi}}]{Shankar:2016}
{Shankar}, F., {Bernardi}, M., {Sheth}, R.~K., {et~al.} 2016, \mnras, 460, 3119

\bibitem[{{Shirasaki} {et~al.}(2016){Shirasaki}, {Komiya}, {Ohishi}, \& {Mizumoto}}]{Shirasaki:2016}
{Shirasaki}, Y., {Komiya}, Y., {Ohishi}, M., \& {Mizumoto}, Y. 2016, \pasj, 68, 23

\bibitem[{Silverman {et~al.}(2009)Silverman, Kovac, Knobel, Lilly, Bolzonella, Lamareille, Mainieri, Brusa, Cappelluti, Peng, Hasinger, Zamorani, Scodeggio, Contini, Carollo, Jahnke, Kneib, Fevre, Bardelli, Bongiorno, Brunner, Caputi, Civano, Comastri, Coppa, Cucciati, de~la Torre, de~Ravel, Elvis, Finoguenov, Fiore, Franzetti, Garilli, Gilli, Griffiths, Iovino, Kampczyk, Koekemoer, Borgne, Brun, Maier, Mignoli, Pello, Montero, Ricciardelli, Tanaka, Tasca, Tresse, Vergani, Vignali, Zucca, Bottini, Cappi, Cassata, Marinoni, McCracken, Memeo, Meneux, Oesch, Porciani, \& Salvato}]{Silverman:2009}
Silverman, J.~D., Kovac, K., Knobel, C., {et~al.} 2009, The Astrophysical Journal, 695, 171

\bibitem[{Singh {et~al.}(2023)Singh, Park, Choi, Kim, Jun, Gibson, Kim, Lee, \& Snaith}]{Singh:2023}
Singh, A., Park, C., Choi, E., {et~al.} 2023, The Astrophysical Journal, 953, 64

\bibitem[{{Siudek} {et~al.}(2023){Siudek}, {Mezcua}, \& {Krywult}}]{Siudek:2023}
{Siudek}, M., {Mezcua}, M., \& {Krywult}, J. 2023, \mnras, 518, 724

\bibitem[{{Viitanen} {et~al.}(2021){Viitanen}, {Allevato}, {Finoguenov}, {Shankar}, \& {Marsden}}]{Viitanen:2021}
{Viitanen}, A., {Allevato}, V., {Finoguenov}, A., {Shankar}, F., \& {Marsden}, C. 2021, \mnras, 507, 6148

\bibitem[{{Wang} {et~al.}(2022){Wang}, {Banerjee}, \& {Abel}}]{Wang_2022}
{Wang}, Y., {Banerjee}, A., \& {Abel}, T. 2022, \mnras, 514, 3828

\bibitem[{{Yang} {et~al.}(2018){Yang}, {Brandt}, {Darvish}, {Chen}, {Vito}, {Alexander}, {Bauer}, \& {Trump}}]{Yang:2018}
{Yang}, G., {Brandt}, W.~N., {Darvish}, B., {et~al.} 2018, \mnras, 480, 1022

\bibitem[{Yuan {et~al.}(2023)Yuan, Zamora, \& Abel}]{Yuan_2023}
Yuan, S., Zamora, A., \& Abel, T. 2023, Monthly Notices of the Royal Astronomical Society, 522, 3935–3947

\end{thebibliography}
\bibliographystyle{aa}

\appendix

\section{Effects of Flux Limited 2MRS Galaxy Sample on $k$NN Distances}
\label{sec:appendixA}

The 2MRS galaxy sample used in this study is flux-limited, which introduces certain effects on the $k$NN distance measurements. Specifically, as redshift increases, the survey completeness becomes increasingly limited to only the most luminous (and massive) galaxies, resulting in a lower overall number density. There are $\sim 20$k galaxies in the low-$z$ range ($0.01 < z < 0.03$) versus $\sim 17$k in the higher-$z$ range ($0.03 < z < 0.06$), corresponding to a$\sim 88\%$ decrease in spatial density. 

\begin{figure}
    \centering \includegraphics[width=\linewidth]{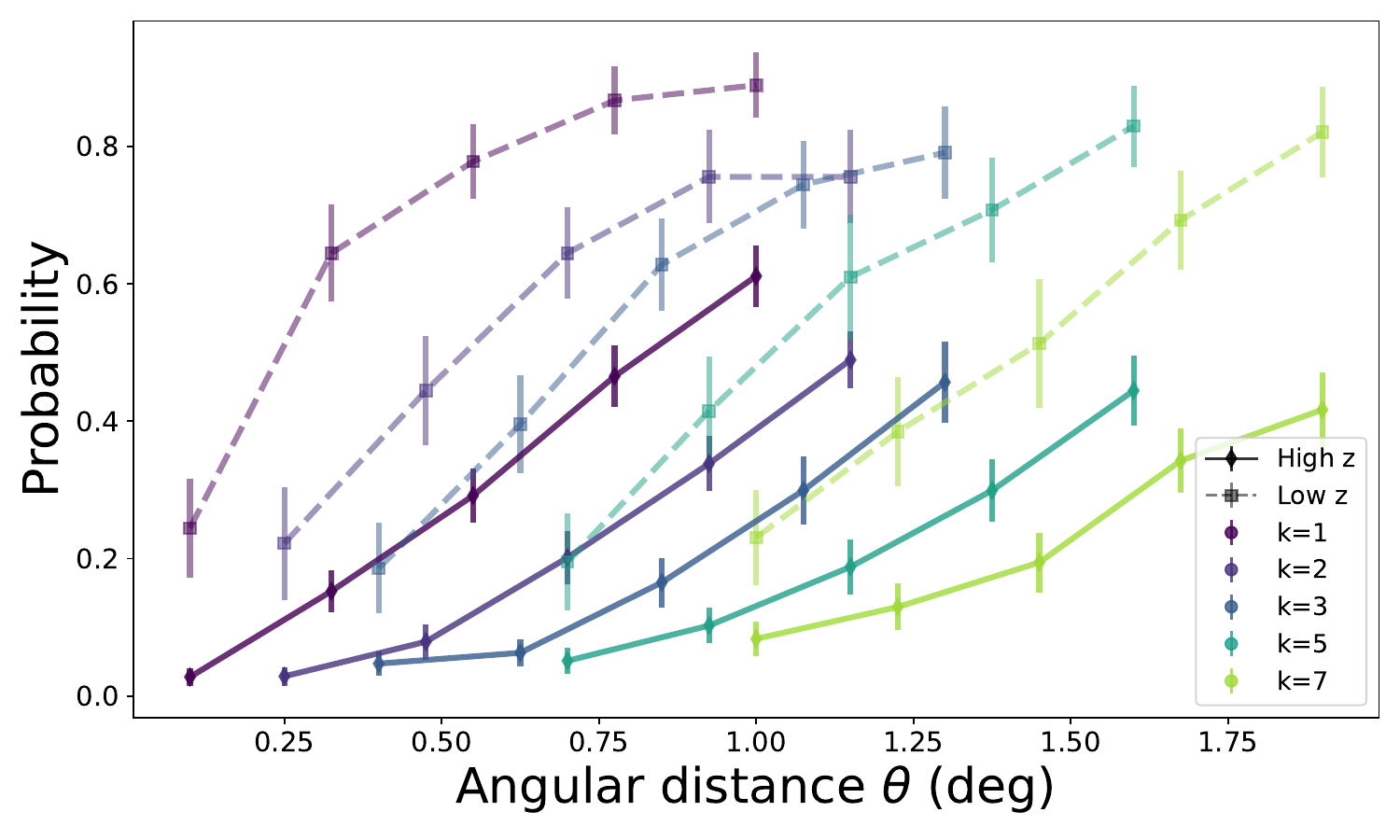}
    \includegraphics[width=\linewidth]{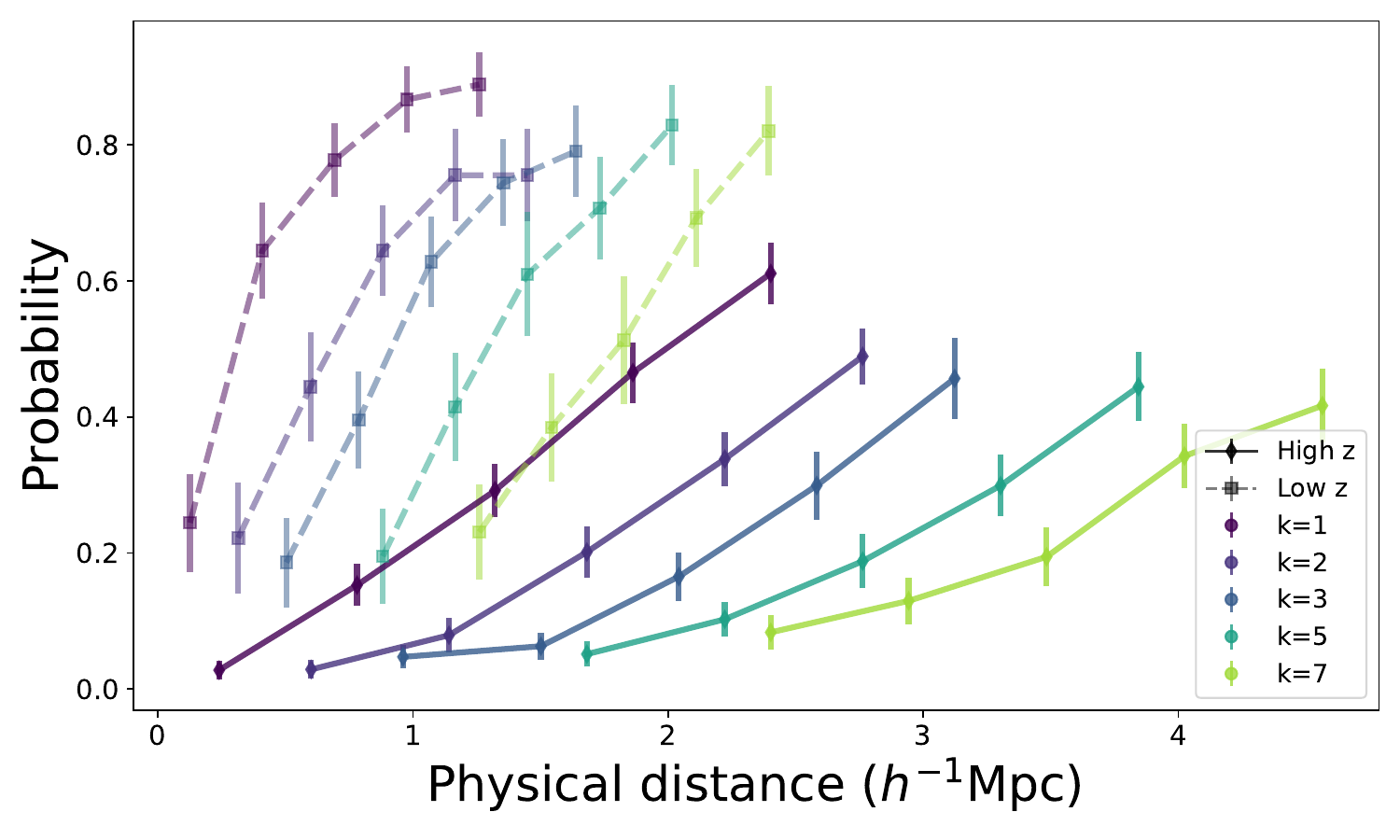}    \caption{Cumulative distribution functions (CDFs) of the $k$NN distances for a volume-limited AGN sample in the low-redshift range ($0.01 < z < 0.03$, dashed lines) and higher-redshift range ($0.03 < z < 0.06$, solid lines). The top plot shows the CDFs as a function of angular distance, while the bottom plot shows the corresponding physical projected separation assuming the average redshift of each $z-$bin. AGN in the lower redshift range have closer galactic neighbors than AGN in the higher redshift range due to the 2MRS flux limit.}
    \label{fig:rshiftcut}
\end{figure}

To show the difference of the $k$NN distances for each redshift range due only to the 2MRS flux limit, we select a volume-limited sample of AGN ($\log L_X>44.5$ [erg/s]) and compute the angular distances to the first 7 neighbors. 
Figure \ref{fig:rshiftcut} shows that AGN in the lower redshift range tend to have systematically closer neighbors than those in the higher redshift range. This supports the idea that, due to the 2MRS flux limit, the two redshift ranges analyzed in this study ($0.01 < z < 0.03$ and $0.03 < z < 0.06$) effectively probe different physical scales. In the lower redshift range, the higher number density of galaxies allows for $k$NN measurements on smaller angular scales, while the higher redshift range, with its sparser galaxy sample, extends the neighbor distances to larger scales.

\section{Random Subsample Splits}
\label{sec:appendixB}

\begin{figure}
    \centering
    \includegraphics[width=\linewidth]{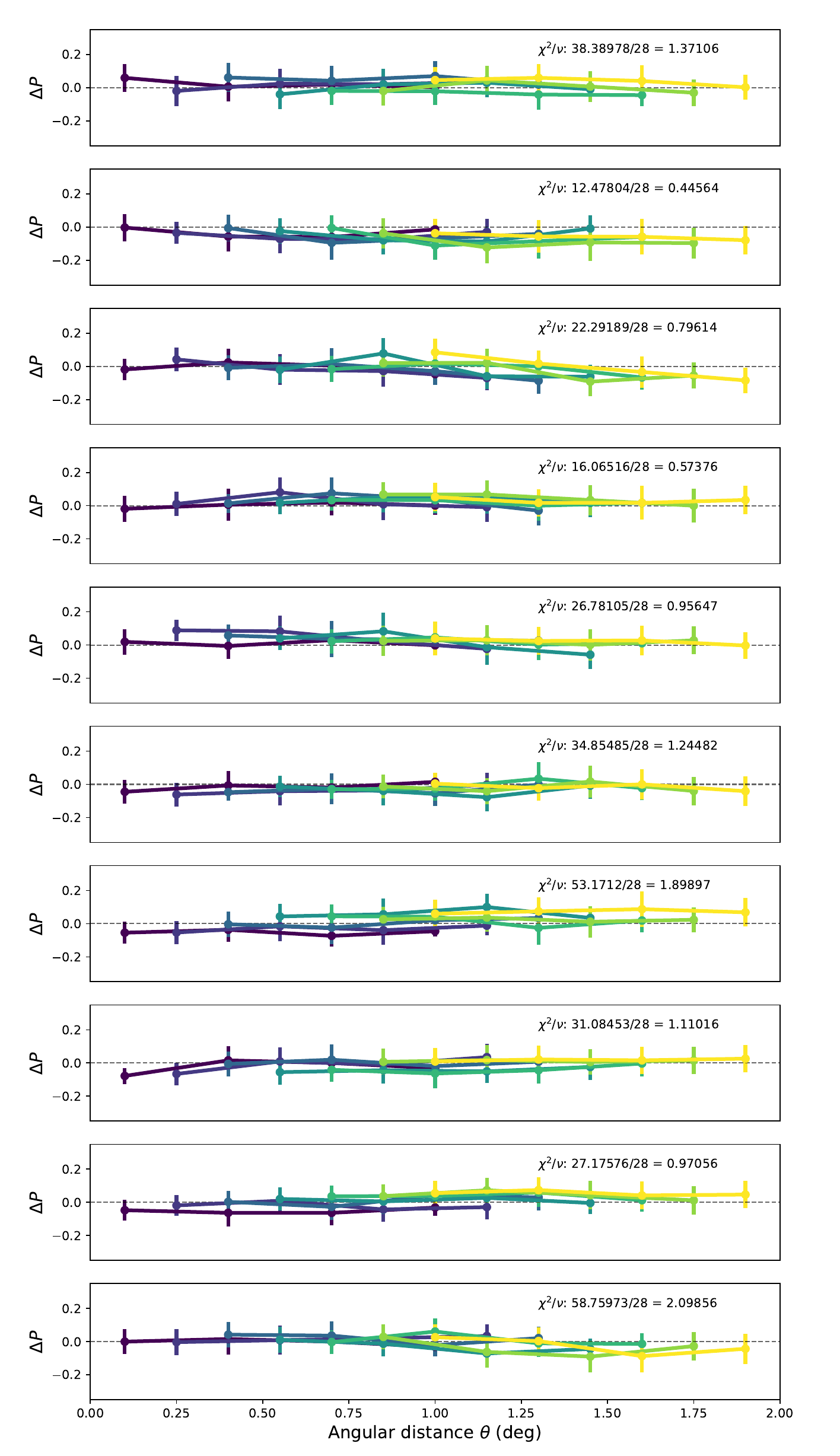}
    \caption{$k$NN CDF differences vs. angular scale of randomly-selected subsamples of the low-redshift BASS AGN ($0.01<z<0.03$). Each subsample contains 68 AGN to roughly match the number in each bin of AGN property. }
    \label{fig:rsplit}
\end{figure}

As discussed in Section \ref{sec:methods}, we split the low-redshift sample ($0.01<z<0.03$) into multiple random subsamples to ensure that observed trends were not driven by sample variance. Each subsample contained 68 AGN, roughly the number in each parameter bin of our main analysis. 
We show in Fig. \ref{fig:rsplit} 10 examples of the $k$NN differences between the random subsamples. These differences are not significant, and notably smaller than those calculated for the parameter bins.

\section{$k$NN Statistics using consistently-measured $M_{\rm BH}$}
\label{sec:appendixC}

    \begin{figure*}
      \centering
      \includegraphics[width=0.49\textwidth]{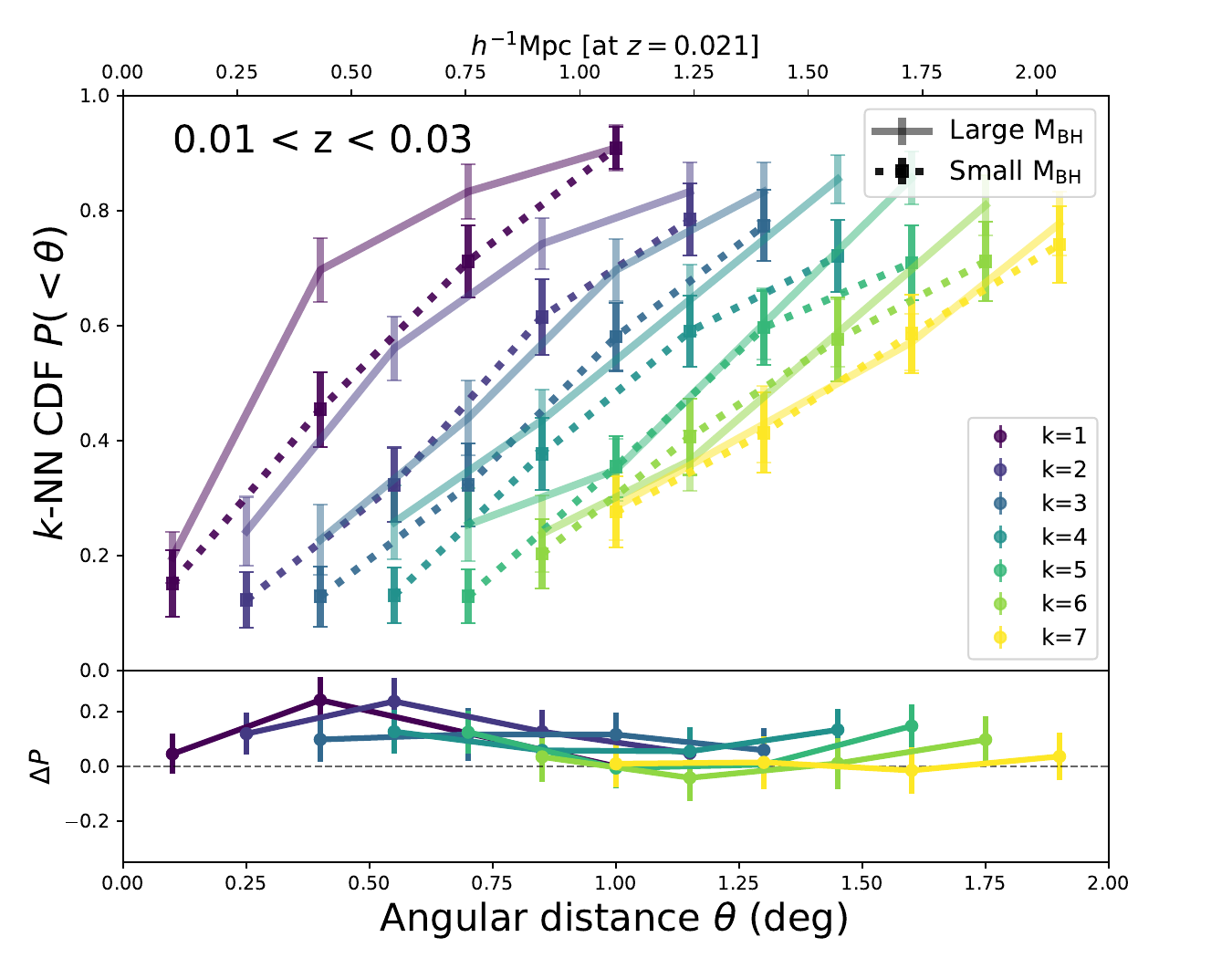}
      \includegraphics[width=0.49\textwidth]{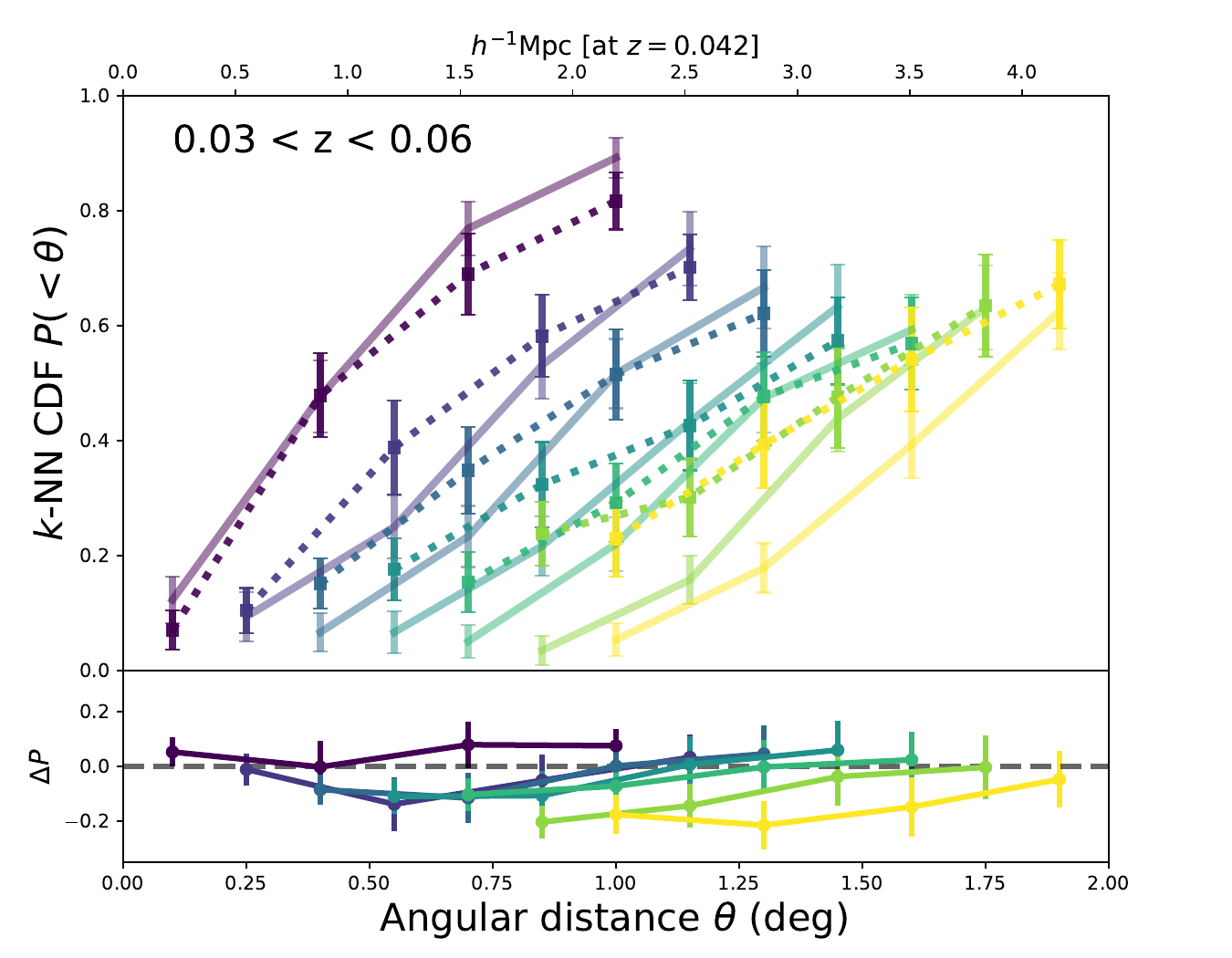}
      \includegraphics[width=0.49\textwidth]{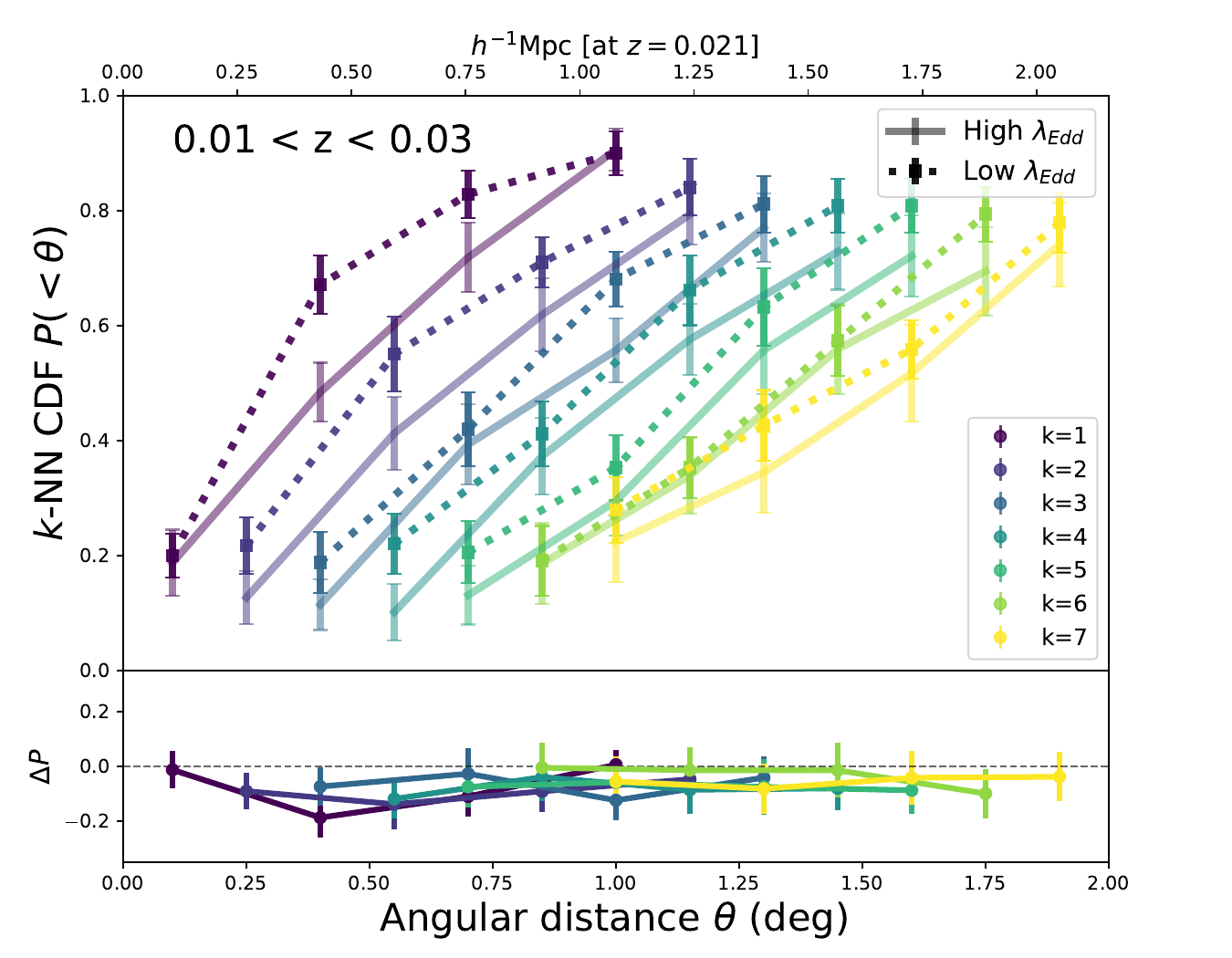}
      \includegraphics[width=0.49\textwidth]{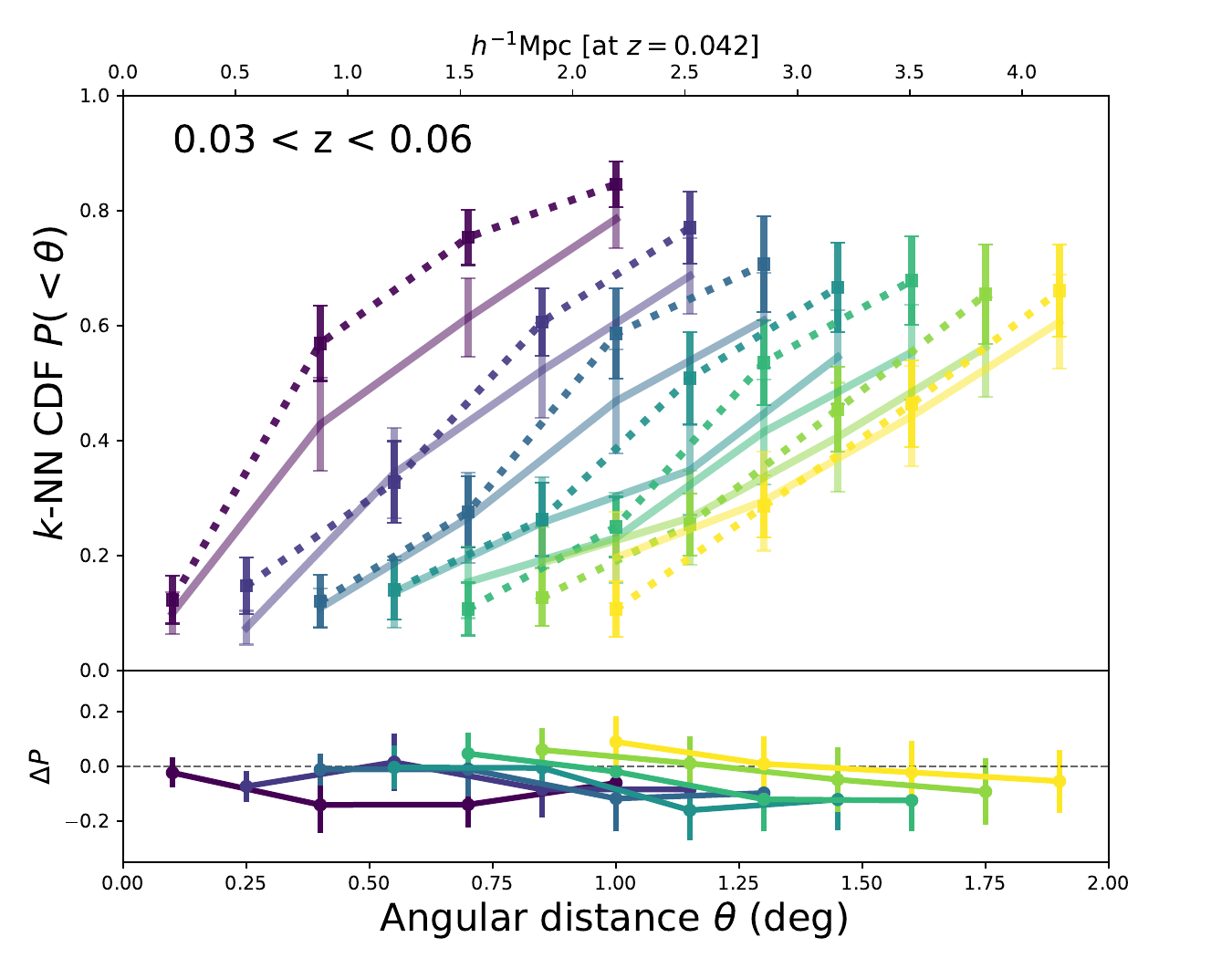}
      \caption{$k$NN CDFs as a function of black hole mass (top) and Eddington ratio (bottom), using values derived from stellar velocity dispersion measurements \citep{caglar:2023}. Left plots show results for AGN in the low-redshift range, and right panels show the higher redshift range. The color scheme is the same as in Figs. 4, 5 and 6, and differences are again shown in the bottom panels.}
      \label{fig:calgar-CDF}
    \end{figure*}
    
The BASS sample uses several mass measurement techniques due to both Type 1 and Type 2 AGN included in the sample. While broad line measurements available from Type 1 spectra are typically preferred over those estimated from velocity dispersion, \cite{caglar:2023} found that extinction can bias these measurements by a significant margin. Additionally, the \cite{Kormendy:2013} $M_{\rm BH}-\sigma_{*}$ relation used to infer $M_{\rm BH}$ in the Type 2 AGN has been found to be too steep for this sample based on Type 1 velocity dispersion measurements \citep{caglar:2023}.

To test whether these biases affect our $k$NN results, we repeat the measurements using the sample of BASS Type1 and Type 2 AGN with $\sigma_{*}$ measurements. We define $M_{\rm BH}$ and Eddington ratio bins by the values obtained from the velocity dispersions using the $M_{\rm BH}-\sigma_{*}$ as measured for this sample in \citep{Caglar:2020} and the same method as described in Section \ref{sec:data}. In doing so, a consistent measure of $M_{\rm BH}$ is used. Figure \ref{fig:calgar-CDF} shows the $k$NN trends with $M_{\rm BH}$ and $\lambda_{\rm Edd}$ using these alternative values. The $k$NN differences are consistent with those found in Section \ref{sec:results} from Figs. \ref{fig:bh-mass-CDF} and \ref{fig:edd-CDF}.

    \begin{figure*}
      \centering
      \includegraphics[width=0.49\textwidth]{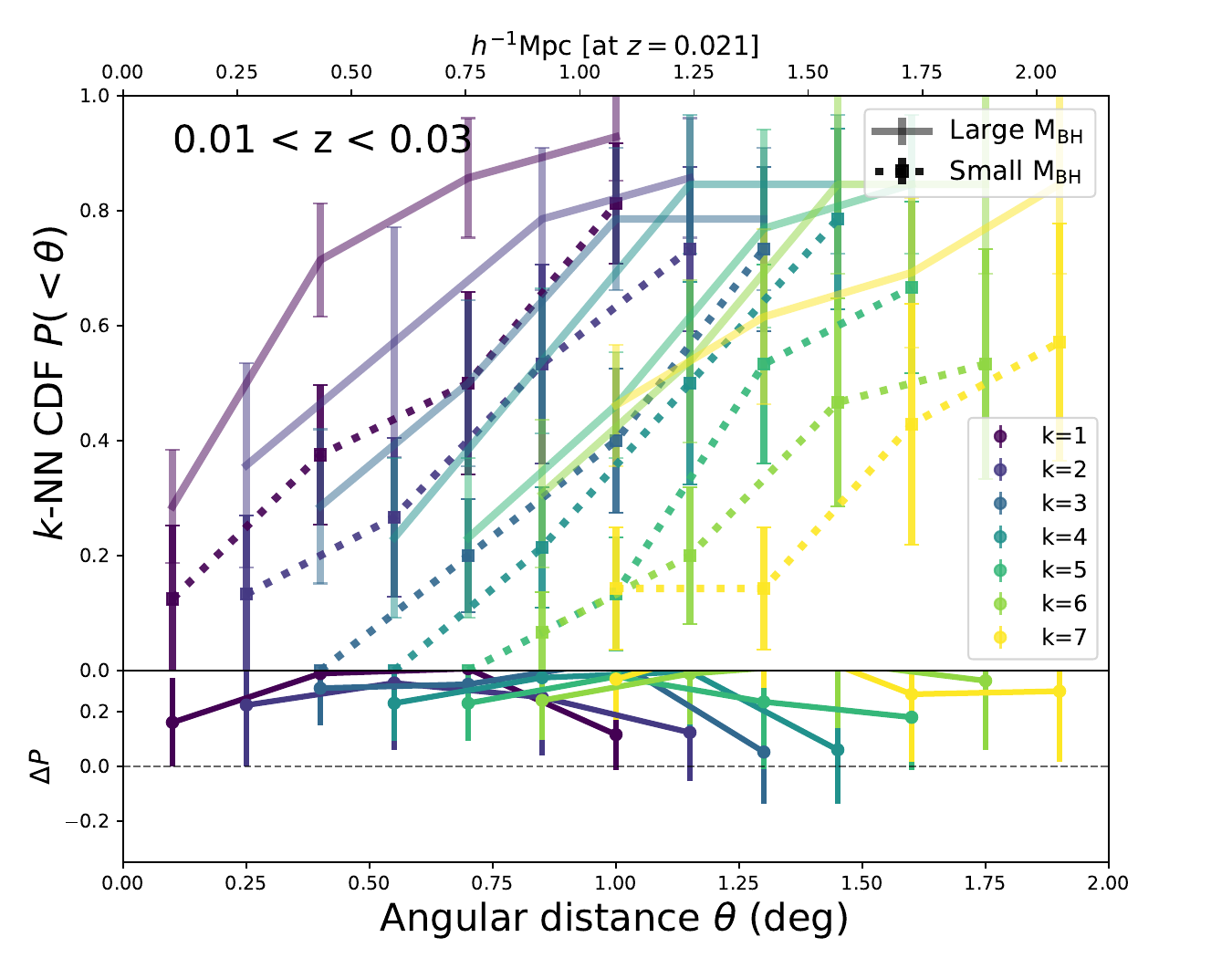}
      \includegraphics[width=0.49\textwidth]{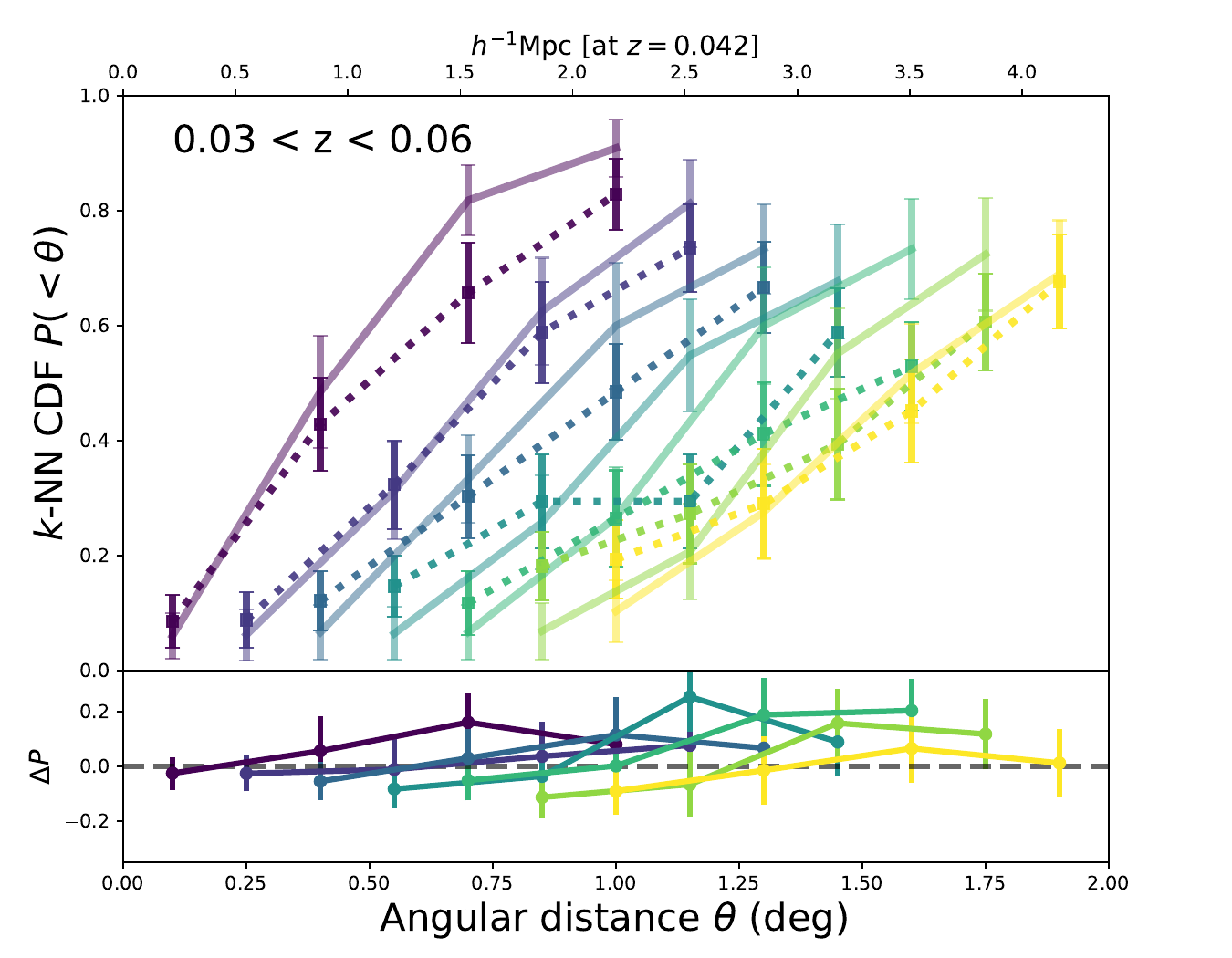}
      \includegraphics[width=0.49\textwidth]{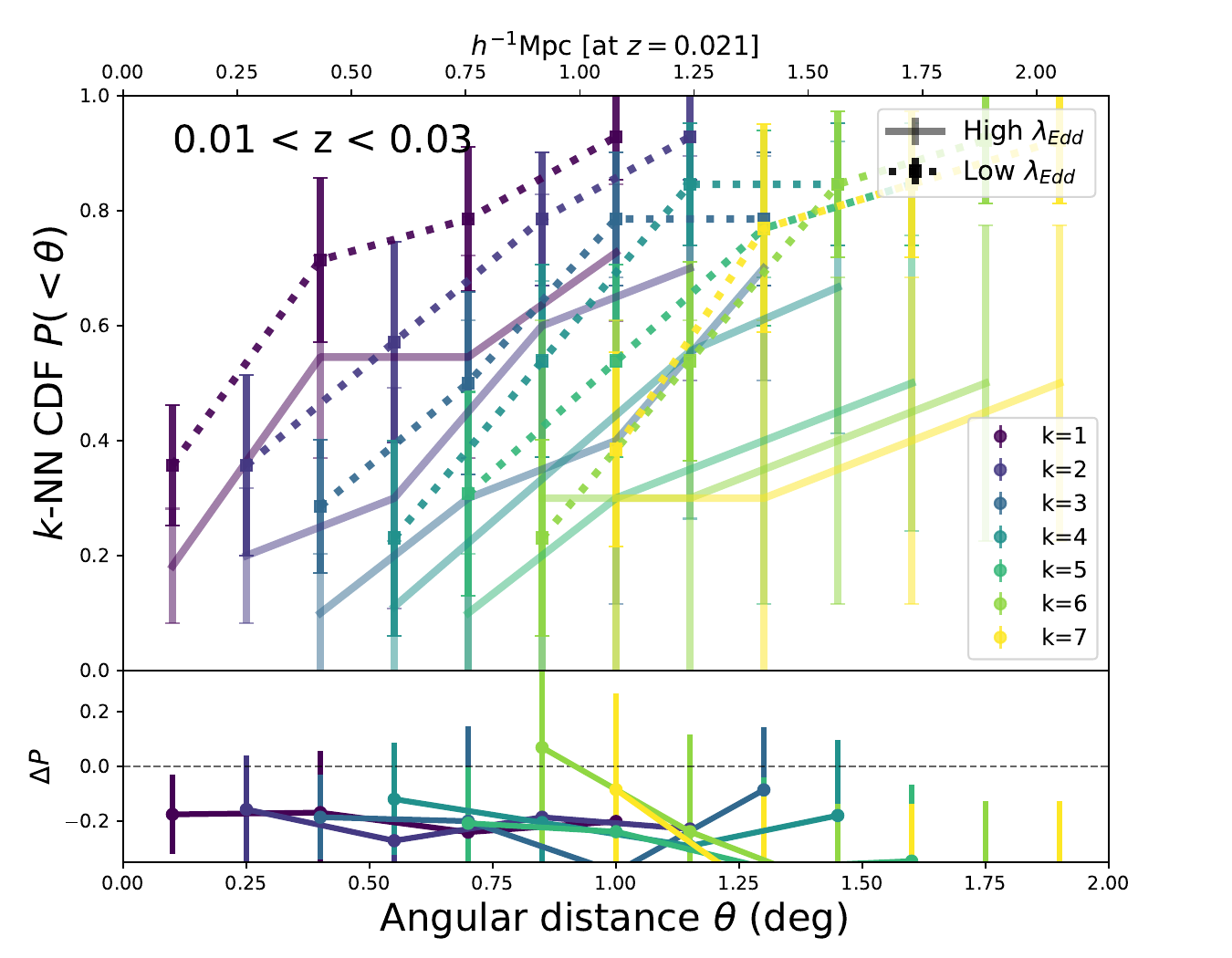}
      \includegraphics[width=0.49\textwidth]{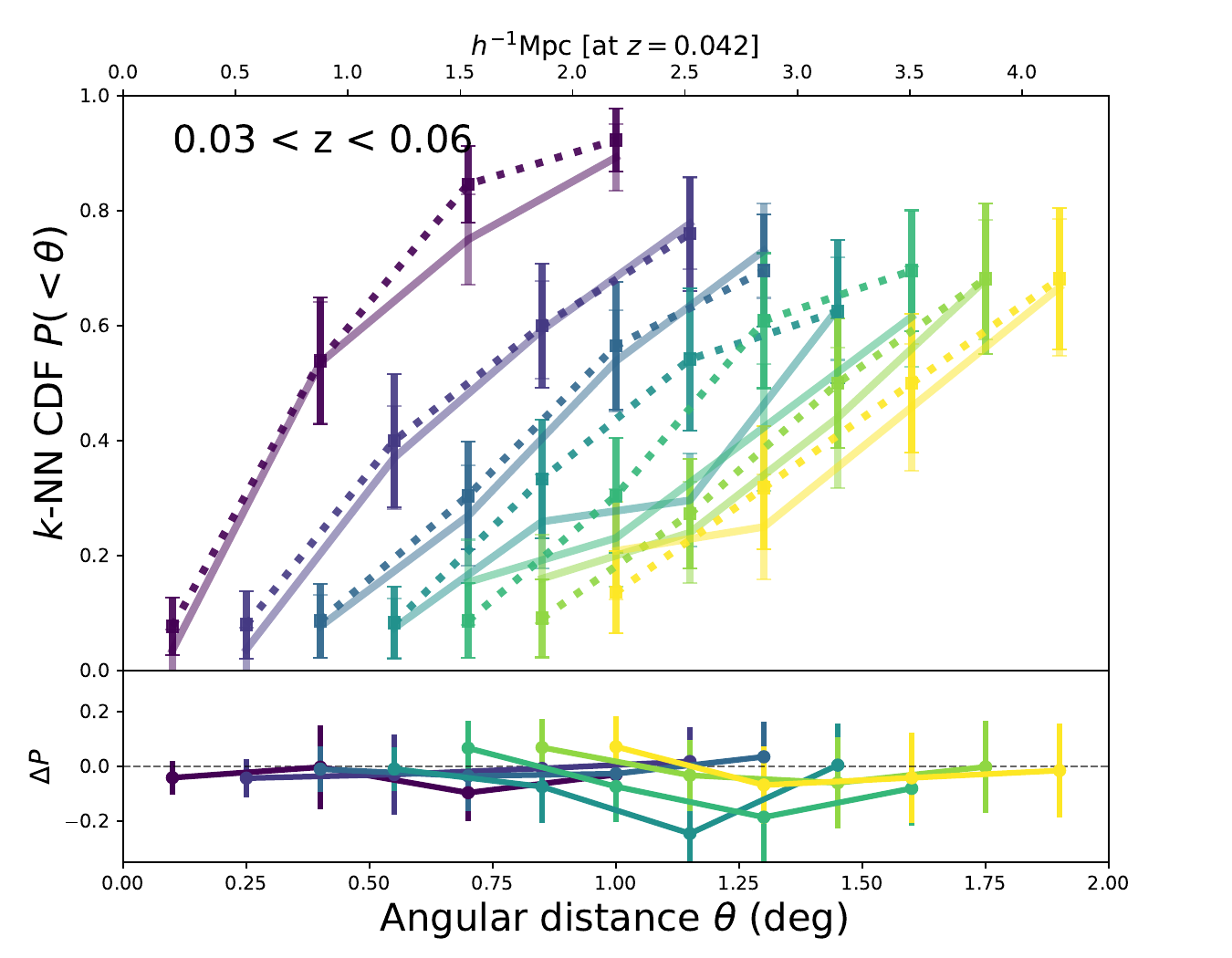}
      \caption{$k$NN CDFs as a function of black hole mass (top) and Eddington ratio (bottom), using values derived from the H$\alpha$ or H$\beta$ broad lines. Left plots show results for AGN in the low-redshift range, and right panels show the higher redshift range. The color scheme is the same as in Figs. 4, 5 and 6, and differences are again shown in the bottom panels.}
      \label{fig:type1-CDF}
    \end{figure*}

Finally, we measured the trends using only Type 1 AGN with broad line $M_{B\rm H}$ and again found consistent results, shown in Fig. \ref{fig:type1-CDF}. Despite the measurements being noisier due to the smaller sample, the massive black holes still have significantly closer neighbors than the less-massive bin. We conclude that mass measurement biases are not causing the $k$NN trends we observe.

\end{document}